\documentclass[notitlepage,english,aps,prl,twocolumn,10pt,tightenlines,floatfix,longbibliography,superscriptaddress]{revtex4-1}

\usepackage{graphicx}
\usepackage{dcolumn}
\usepackage{bm}
\usepackage{color}
\usepackage[usenames,dvipsnames]{xcolor}
\usepackage[colorlinks=true,urlcolor=blue,citecolor=blue,linkcolor=blue]{hyperref}
\usepackage{amssymb,ulem,amsmath}

\usepackage{siunitx}
\newcommand{\ket}[1]{\left|#1\right\rangle}
\newcommand{\braket}[2]{\left\langle #1 | #2 \right\rangle}
\newcommand{\qql}{\textquotedblleft}
\newcommand{\qqr}{\textquotedblright}
\newcommand{\vc}[1]{\bm{\mathrm{#1}}}

\begin{document}

\title{Geometric origin of superfluidity in the Lieb lattice flat band}

\author{Aleksi Julku}

\author{Sebastiano Peotta}

\author{Tuomas I. Vanhala}

\affiliation{COMP Centre of Excellence, Department of Applied Physics, Aalto University School of Science, FI-00076 Aalto, Finland}

\author{Dong-Hee Kim}

\affiliation{Department of Physics and Photon Science, School of Physics and Chemistry, Gwangju Institute of Science and Technology, Gwangju 61005, Korea}

\author{P\"aivi T\"orm\"a}

\affiliation{COMP Centre of Excellence, Department of Applied Physics, Aalto University School of Science, FI-00076 Aalto, Finland}

\begin{abstract}
The ground state and transport properties of the Lieb lattice flat band in the presence of an attractive Hubbard interaction are considered. It is shown that the superfluid weight can  be large even for an isolated and strictly flat band. Moreover the superfluid weight is proportional to the interaction strength and to the quantum metric, a band structure quantity derived solely from the flat-band Bloch functions. These predictions are amenable to verification with ultracold gases and may explain the anomalous behaviour of the superfluid weight of high-$T_{\rm c}$ superconductors.
\end{abstract}

\maketitle

A flat band is a Bloch band with constant energy dispersion $\varepsilon_{n\vc{k}} \approx \varepsilon_n$ ($n$ is the band index) as a function of quasi-momentum $\vc{k}$ and is composed of localized eigenstates. In absence of disorder and interactions the ground state of a flat band is insulating at any filling~\cite{Resta:2011}. However, interactions and disorder lead to a reconstruction of the ground state whose properties are often hard to predict. Bands that are nearly flat and/or feature nontrivial topological invariant, similar to Landau levels producing the quantum Hall effects~\cite{QHE1,QHE2,QHE3}, have been considered in recent theoretical works~\cite{Huber:2010,Sun:2011,Tang:2011,Neupert:2011,Bregholtz:2013,Roy:2014,Takayoshi:2013,Tovmasyan:2013} and can be realized in ultracold gas experiments~\cite{Aidelsburger:2013,Miyake:2013,Aidelsburger:2015}. Flat-band ferromagnetism has been studied first by Lieb~\cite{Lieb:1989} and, subsequently, by Tasaki and Mielke~\cite{Mielke:1991,Tasaki:1992,Mielke:1993,Tasaki:1998}. More recently it has been shown that the high density of states of flat bands enhances the superconducting critical temperature~\cite{Kopnin:2011,Heikkila:2011}. Indeed, for fixed interaction strength, the flat-band dispersion provides the maximal critical temperature within mean-field BCS theory~\cite{Noda:2015a}.

\begin{figure}
\includegraphics[width=0.9\columnwidth]{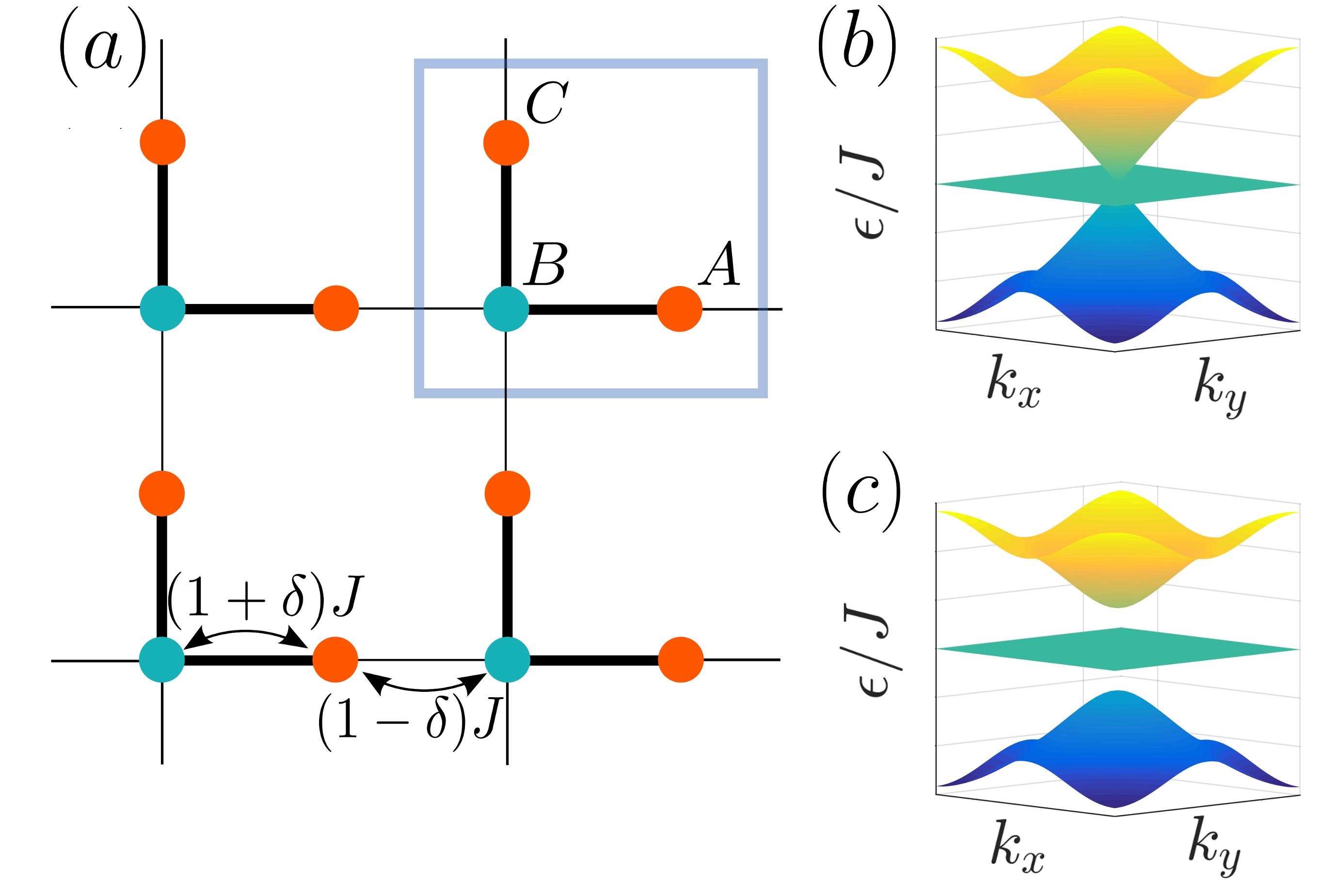}
\caption{(a) The Lieb lattice and its unit cell (grey box) are shown. The orbitals in the unit cell are labelled by $\alpha = A,B,C$. The thick lines represent nearest-neighbour hoppings with energy $(1+\delta)J$, while the hopping energy corresponding to the thin lines is $(1-\delta)J$ with $0 \leq \delta \leq 1$ parametrizing the staggered hopping. (b)-(c) The energy dispersion as a function of quasimomentum $\vc{k}$ for $\delta=0$ (b) and $\delta=0.3$ (c), respectively. The middle band is strictly flat $\varepsilon_{0\vc{k}} = 0$ for any value of $\delta$ while the upper and lower band have dispersions $\varepsilon_{\pm,\vc{k}} = \pm 2J\sqrt{1+\delta^2 +({1-\delta^2})(\cos{k_x a} + \cos{k_y a})/2}$.}
\label{fig:one}
\end{figure}

Flat bands, or quasi-flat bands,  can be realized in bipartite lattices~\cite{Lieb:1989} and other models~\cite{Tasaki:1998,Weeks:2010,Sun:2011,Tang:2011,Neupert:2011}. A simple bipartite lattice featuring a strictly flat band is the Lieb lattice [Fig.~\ref{fig:one}(a)]. Recent  studies on models defined on the Lieb lattice focus on the ferromagnetic and topological properties~\cite{Noda:2009,Goldman:2011,Noda:2014,Noda:2015,Tsai:2015,Palumbo:2015,Dauphin:2016}, while superconductivity has been studied in Refs.~\cite{Noda:2015,Iglovikov:2014}. On the experimental side, a highly tunable Lieb lattice has been realized with ultracold gases~\cite{Taie:2015}. Intriguingly,  the ${\rm CuO}_{2}$ planes responsible for the exotic properties of high-$T_{\rm c}$ cuprate superconductors have the Lieb lattice structure. Thus a Hubbard model on the Lieb lattice~\cite{Mattheiss:1987,Varma:1987,Emery:1987} is a natural, and possibly indispensable~\cite{Kung:2016,Valkov:2016,
Adolphs:2016}, extension of the single-band Hubbard model more commonly used~\cite{LeBlanc:2015}. 

The important question of whether an isolated strictly flat band can support superfluid transport is open. Its answer is of interest for ongoing ultracold gas experiments and may have important implications for the theory of superconductivity. Meissner effect and dissipationless transport are manifestations of a finite superfluid weight that in conventional superconductors at zero temperature reads $D_{s} = n_{\rm p}/m_{\rm eff}$, with $n_{\rm p}$ the particle density and $m_{\rm eff}$ the band effective mass. Interestingly, the superfluid weight of a flat band is not necessarily vanishing, as suggested by $m_{\rm eff}\to +\infty$, but proportional to  the quantum metric~\cite{Peotta:2015}. Flat bands with nonzero Chern number $C$ (the topological index of Landau levels) have nonzero superfluid weight due to the bound $D_{\rm s} \geq |C|$. For a large class of Hamiltonians defined on the Lieb lattice the flat band has $C=0$~\cite{Chen:2014}. Lower bounds on $D_{\rm s}$ are not available at present for topologically trivial bands or bands characterized by other topological invariants than the Chern number.

Here we consider  a tight-binding model with attractive Hubbard interaction on the Lieb lattice.
This model features a strictly flat band with $C=0$. We show that the total superfluid weight tensor receives contributions from the flat band, $\left.D_{\rm s}\right|_{\rm f.b.}$, and from the other bands, $\left.D_{\rm s}\right|_{\rm o.b.}$, that is,
$D_{\rm s} = \left.D_{\rm s}\right|_{\rm f.b.}+\left.D_{\rm s}\right|_{\rm o.b.}$.
We find that $\left.D_{\rm s}\right|_{\rm f.b.}$ depends on the flat-band Bloch functions through the quantum metric. This is called a \qql geometric\qqr contribution distinct from the \qql conventional\qqr contribution, which depends only on the derivatives of $\varepsilon_{n\vc{k}}$~\cite{Peotta:2015}. Only the latter is accounted for when evaluating the superfluid weight of known superconductors~\cite{Prozorov:2006,Prozorov:2011}.
Importantly, the energy scale of the geometric contribution is the  coupling constant $U$, at odds with the conventional result $D_{s} = n_{\rm p}/m_{\rm eff}\propto J$, where $J$  is the characteristic hopping energy in a tight-binding Hamiltonian. We identify the regimes where $\left.D_{\rm s}\right|_{\rm f.b.}$ dominates over the term  $\left.D_{\rm s}\right|_{\rm o.b.}$, which includes the conventional and geometric contributions of other bands. These results are obtained with mean-field BCS theory. The validity of BCS theory  is rigorously  justified by showing that, in the isolated flat-band limit, the BCS wavefunction is exact for any bipartite lattice. Furthermore we compare the BCS predictions for the pairing order parameters and the superfluid weight, respectively, with dynamical mean-field theory (DMFT) and exact diagonalization (ED), finding good agreement even when the flat band is not isolated.

\noindent\textit{Hubbard model on the Lieb lattice ---}
The Hamiltonian $\mathcal{\hat{H}} = \mathcal{\hat{H}}_{\rm kin} + \mathcal{\hat{H}}_{\rm int}-\mu \hat{N}$ defined on the Lieb lattice  comprises the chemical potential term $-\mu {\hat N}$ ($\hat N$ is the particle number operator), the attractive Hubbard interaction $\mathcal{\hat H}_{\rm int}$ defined below and the kinetic term  $\mathcal{\hat{H}}_{\rm kin} = \sum_{\vc{k},\sigma}\hat{\vc{c}}_{\vc{k}\sigma}^\dagger H_{\vc{k}}\hat{\vc{c}}_{\vc{k}\sigma}$ with staggered nearest-neighbour hopping [Fig.~\ref{fig:one}(a)]
\begin{align}
\label{fourierhamiltonian}
H_{\vc{k}}  = 2J 
\begin{pmatrix}
0 & a_{\vc{k}} & 0 \\
a_{\vc{k}}^* & 0 &b_{\vc{k}}  \\
0  & b_{\vc{k}}^* & 0
\end{pmatrix}\,,
\end{align}
where $a_{\vc{k}} = \cos\frac{k_xa}{2} + i\delta\sin\frac{k_x a}{2}$, $b_{\vc{k}} = \cos\frac{k_ya}{2} + i\delta\sin\frac{k_y a}{2}$ and $a$ the lattice constant. The fermion operators are defined as $\hat{\vc{c}}_{\vc{k}\sigma} = (\hat{c}_{A\vc{k}\sigma},\hat{c}_{B\vc{k}\sigma},\hat{c}_{C\vc{k}\sigma})^T$ and $\hat{c}_{\alpha\vc{k}\sigma} = \frac{1}{\sqrt{N_c}} \sum_{\vc{i}} e^{-i\vc{k}\cdot \vc{r}_{\vc{i}\alpha}}\hat{c}_{\vc{i}\alpha\sigma}$  where $N_c$ is the number of unit cells, $\vc{r}_{\vc{i}\alpha}$ is the position vector of the $\alpha$ orbital in the $\vc{i}$-th unit cell [$\,\vc{i} = (i_x,i_y)^T\,$] and the operator $\hat{c}_{\vc{i}\alpha\sigma}$ annihilates a fermion with spin $\sigma=\uparrow,\downarrow$ in the orbital centered at $\vc{r}_{\vc{i}\alpha}$. By solving the eigenvalue problem $H_{\vc{k}}|g_{n\vc{k}}\rangle = \varepsilon_{n\vc{k}} |g_{n\vc{k}}\rangle$ one obtains the Bloch functions $|g_{n\vc{k}}\rangle$ and the band dispersions $\varepsilon_{n\vc{k}}$ ($n = 0,\pm$).
The middle band is strictly flat ($\varepsilon_{n = 0,\vc{k}} = 0$) for any value of the staggered-hopping parameter $\delta$ and isolated from the other bands by an energy gap $E_{\rm gap} = \sqrt{8}J \delta$. As in Ref.~\cite{Iglovikov:2014},  the interaction term $\mathcal{\hat H}_{\rm int}= -U\sum_{\vc{i},\alpha}(\hat{n}_{\vc{i}\alpha \uparrow}-1/2)(\hat{n}_{\vc{i}\alpha \downarrow}-1/2)$, where $U >0$ and $\hat{n}_{\vc{i}\alpha \sigma} = \hat{c}^\dag_{\vc{i}\alpha \sigma}\hat{c}_{\vc{i}\alpha \sigma}$, is approximated by mean-field pairing $\Delta_{\alpha} = -U \langle \hat{c}_{\vc{i}\alpha\downarrow}\hat{c}_{\vc{i}\alpha\uparrow}\rangle$ and Hartree potentials $n_{\alpha} = \langle \hat{n}_{\vc{i}\alpha\sigma}\rangle$
\begin{equation}
\label{interaction_ham}
\begin{split}
\hat{\mathcal{H}}_{\rm int} \approx &\sum_{\vc{i},\alpha}\left( \Delta_{\alpha}\hat{c}_{\vc{i}\alpha\uparrow}^\dagger\hat{c}_{\vc{i}\alpha\downarrow}^\dagger + {\rm H.c.}\right) +U\sum_{\vc{i},\alpha,\sigma}\left(n_\alpha -\frac{1}{2}\right)\hat{n}_{\vc{i}\alpha \sigma}\,.
\end{split}
\end{equation}
The equivalence of orbitals $A$ and $C$ implies  $\Delta_A = \Delta_C$ and $n_{A} = n_{C}$. From the zero-temperature gap equations at half-filling $\nu = \sum_\alpha n_\alpha = 3/2$ one finds $\Delta_{A} = U/4$ and $\Delta_B = 0$  at leading order in $U/J$~\cite{suppl}.

\begin{figure}
  \centering
    \includegraphics[width=1.0\columnwidth]{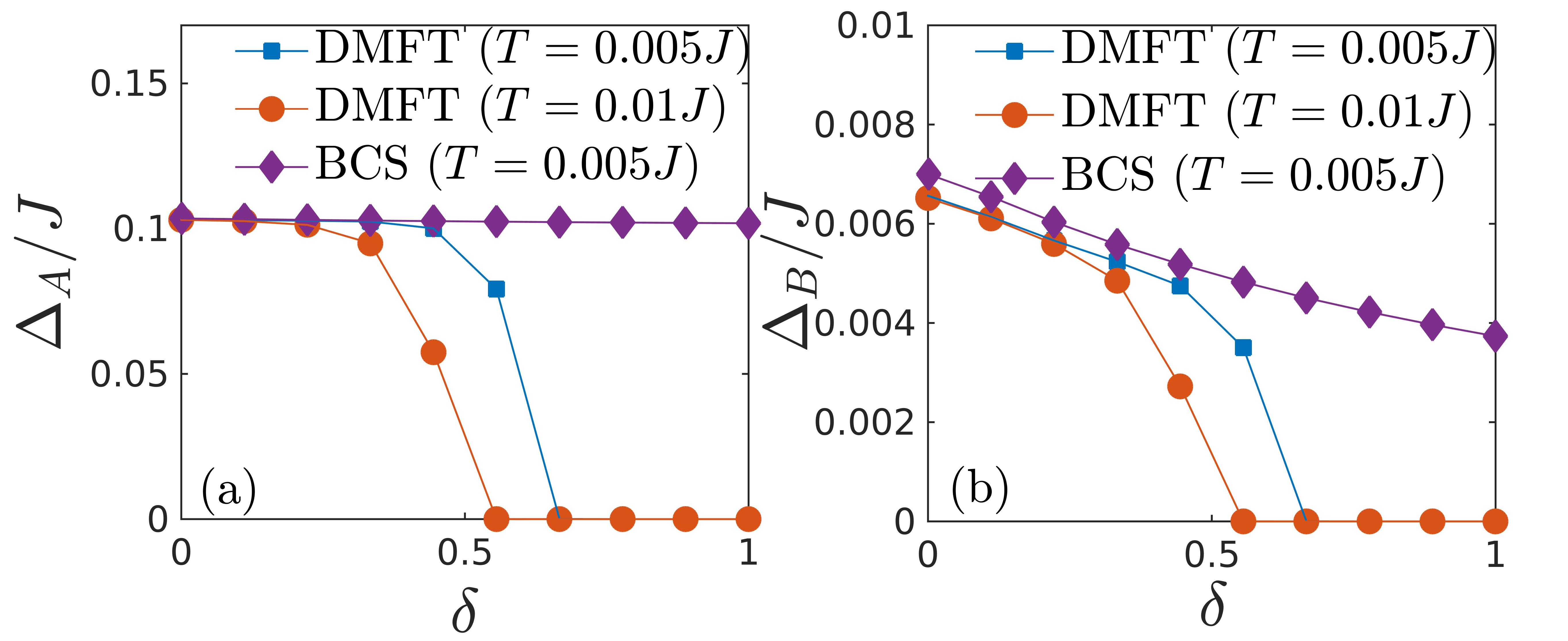}
   \caption{Order parameters $\Delta_A/J$ (left) and $\Delta_B/J$ (right) as a function of $\delta$ obtained with DMFT and mean-field BCS theory at temperatures $k_{\rm B}T = 5\cdot 10^{-3}J$ and $10^{-2}J$, filling $\nu = 1.5$ and coupling strength $U=0.4J$. At these temperatures, significantly lower than the BCS critical temperature $k_{\rm B}T_{\rm c, BCS} \approx \Delta_{\rm A}/2 = U/8 = 5\cdot 10^{-2}J$, the BCS results are indistinguishable from the zero temperatures ones.}
   \label{fig:two}
\end{figure}

\noindent\textit{Exactness of BCS wavefunction for a flat band ---} Lieb theorem~\cite{Lieb:1989} states that the ground state at half-filling of a bipartite lattice with repulsive Hubbard interaction has total spin $S = N_{\rm c}N_{\rm f.b.}/2$, where $N_{\rm f.b.}$ is the number of flat bands and $N_c$ the number unit cells. The Lieb lattice has $N_{\rm f.b.} = 1$ and if $U \ll E_{\rm gap}$, the completely filled lower band can be neglected at half-filling. The ferromagnetic wavefunctions $|\text{Ferro}\rangle = \prod_{\vc{k}}\big(u d^\dag_{0 \vc{k} \downarrow} + v d^\dag_{0 \vc{k} \uparrow} \big)|\emptyset\rangle$, parametrized by $u,v$ with $|u|^2+|v|^2=1$, have total spin $S$ and therefore are the only ground states. Here the operator $d^\dag_{n = 0,\vc{k}\sigma}$ creates a fermion within the flat band. A repulsive Hubbard model on a bipartite lattice can be mapped by a particle-hole transformation into an attractive one~\cite{Emery:1976}. Under this transformation the state $|\text{Ferro}\rangle$ is mapped into a BCS wavefunction $|\text{BCS}\rangle = \prod_{\vc{k}}\big(u + v d^\dag_{0\vc{k}\uparrow} d^\dag_{0(-\vc{k})\downarrow} \big)|\emptyset\rangle$ and the spin operator along the $z-$axis $\hat {S}_{\vc{i}\alpha}^{z} = \frac{1}{2}(\hat{n}_{\vc{i}\alpha\uparrow}-\hat{n}_{\vc{i}\alpha\downarrow})$ into the operator $\hat{\Delta}_{\vc{i}\alpha}^{z} = \frac{1}{2}(\hat{n}_{\vc{i}\alpha\uparrow}+\hat{n}_{\vc{i}\alpha\downarrow}-1)$. The expectation value $\langle \sum_{\alpha}\hat{\Delta}^z_{\vc{i}\alpha}\rangle = \nu -3/2$ gives the filling $\nu$. Therefore the BCS wavefunction is the exact ground state for arbitrary flat band filling. This result is easily extended to any bipartite lattice. Consistently with this result, the numerical data obtained with DMFT and ED converge to the predictions of BCS theory for small $U$ and partially filled flat band, as we show below and in Ref.~\cite{suppl}.

\noindent\textit{Comparison with DMFT ---} To investigate the accuracy of BCS theory also for a non-isolated flat band, we compare it in Fig.~\ref{fig:two} against DMFT with respect to the pairing potentials (order parameters) $\Delta_A$ [Fig.~\ref{fig:two} (a)] and $\Delta_B$ [Fig.~\ref{fig:two} (b)] as a function of $\delta$ at half filling. We use cellular dynamical mean-field theory~\cite{Maier:2005,Vanhala:2015} with continuous-time interaction-expansion impurity solver~\cite{Rubtsov:2005,Gull:2011},
which treats correlations exactly within the three-site unit cell and goes beyond mean-field BCS theory.
For small $\delta$, DMFT is in good agreement with BCS, especially regarding $\Delta_A$. The results for large $\delta$ are discussed below. In particular, both methods show that, even when $\delta = E_{\rm gap} =0$, pairing is dominated by the flat band and the effect of the other bands is small. 

\begin{figure}
  \centering
    \includegraphics[width=1.0\columnwidth]{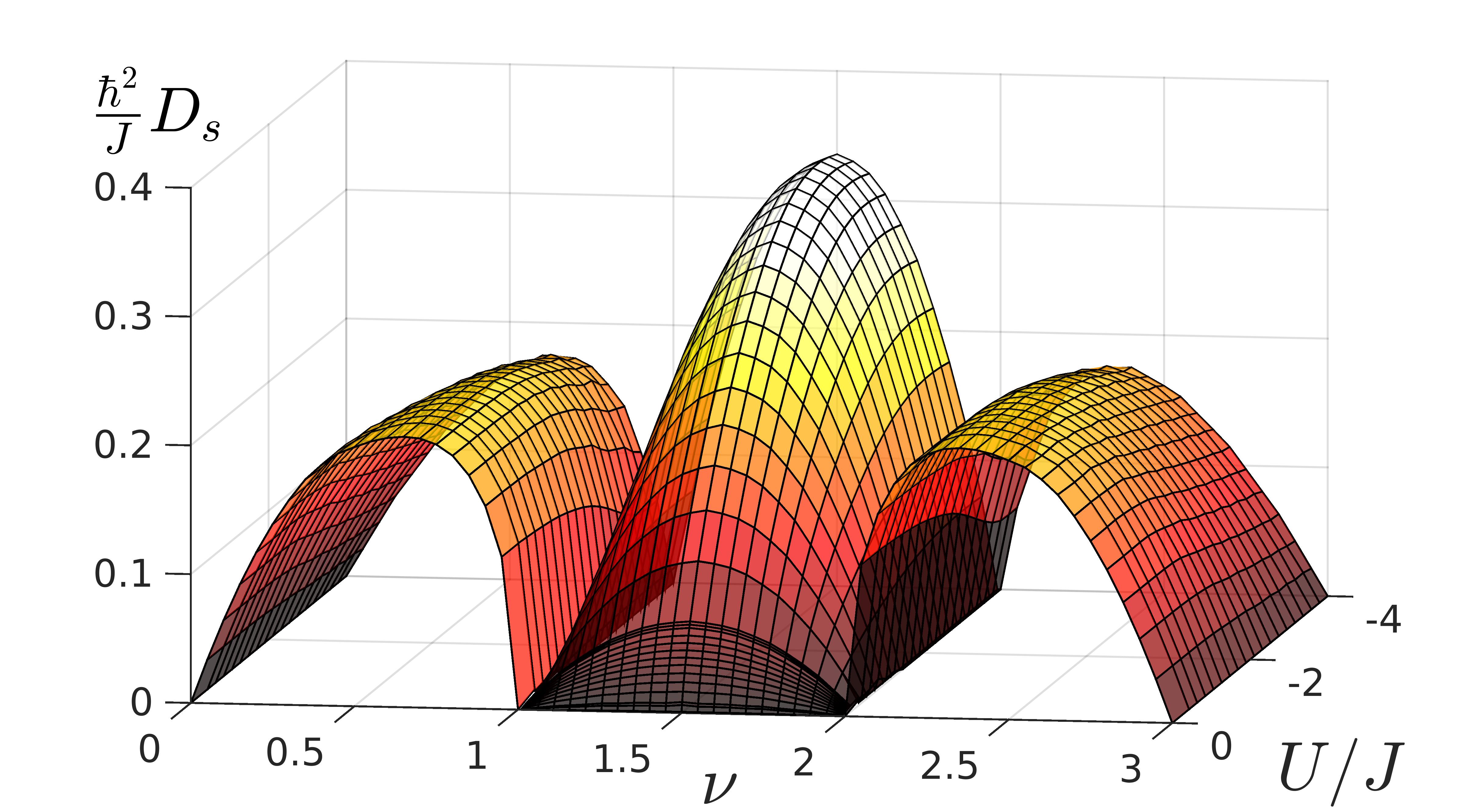}
   \caption{Diagonal components of the superfluid weight tensor $[D_{\rm s}]_{x,x} = [D_{\rm s}]_{y,y}\approx D_{\rm s}$ as a function of interaction $U/J$ and filling $\nu$ for $\delta =10^{-3}$ and at zero temperature. The superfluid weight for partially filled flat band ($1\leq \nu \leq 2$) depends strongly on $U$ in contrast to the other bands.}
  \label{fig:three}
\end{figure}

\begin{figure*}
\includegraphics[width=1.0\textwidth]{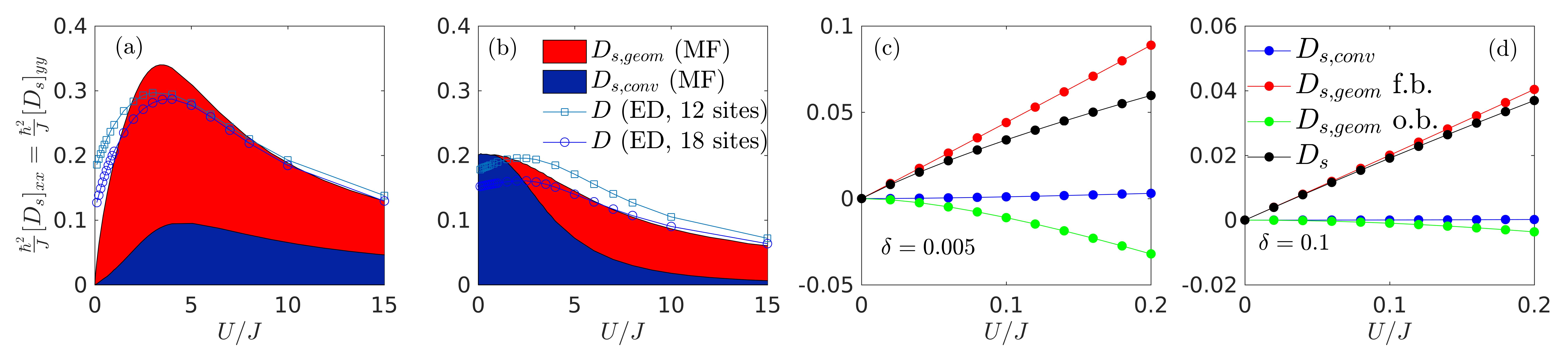}
   \caption{\label{fig:four}(a)-(b) Conventional superfluid weight $D_{\rm s,conv}$ (blue area) compared with the geometric one $D_{\rm s,geom}$ (red area) for $\nu = 1.5$ (a) and $\nu = 2.5$ (b). Here $T=0$ and $\delta = 10^{-3}$. Also the Drude weight $D$ obtained from ED is shown. Squares and circles correspond to the 12 sites and 18 sites clusters, respectively. Data for the 24 sites cluster at $\nu = 2.5$ (shown in~\cite{suppl}) does not deviate significantly with respect to 18 sites. (c)-(d) Various superfluid weight contributions for half-filled flat band, small $U \leq 0.2J$, $\delta =5\cdot 10^{-3}$ (c), $\delta=0.1$ (d).}
\end{figure*}

\noindent\textit{Superfluid weight ---} The superfluid weight is defined as the change in free energy density $\Delta f = \frac{1}{8}D_{\rm s}(\hbar\vc{q})^2$ due to the winding with wavevector $\vc{q}$ of the order parameter phase $\Delta(\vc{r}) = \Delta e^{2i\vc{q}\cdot \vc{r}}$. The superfluid weight obtained from multiband BCS theory is shown in Fig.~\ref{fig:three} as a function of coupling $U$ and filling $\nu$  for zero temperature and $\delta = 10^{-3}$~\cite{suppl}. The Hartree term of Eq.~(\ref{interaction_ham}) is needed for preserving the $SU(2)$ symmetry that allows to calculate $D_{\rm s}$ for arbitrary flat band fillings~\cite{Iglovikov:2014}. This symmetry corresponds, under the  particle-hole transformation, to the spin rotational symmetry of the repulsive Hubbard model. For $\delta\neq 0$, the superfluid weight tensor acquires nonzero off-diagonal components $[D_{\rm s}]_{x,y} = [D_{\rm s}]_{y,x}$.  However, this effect is small and we focus only on the diagonal components $[D_{\rm s}]_{x,x} = [D_{\rm s}]_{y,y} \approx D_{\rm s}$. A striking feature of Fig.~\ref{fig:three} is that, for partially filled dispersive bands, $D_s$ is finite and roughly  constant as a function of $U$, while the superfluid weight within the flat band depends strongly on $U$ and has a nonmonotonic behavior [see also Fig.~\ref{fig:four}(a)].  This is consistent with the fact that superconductivity in the dispersive bands emerges from a metallic state with nonzero Drude weight which is the $U\to 0$ limit of $D_{\rm s}$ at zero temperature~\cite{Scalapino:1992,Scalapino:1993}. On the contrary, superconductivity in the flat band smoothly emerges with increasing $U$ from an insulating state with zero Drude weight. Notably, the superfluid weight of a topologically trivial flat band can be nonzero and larger than the one of dispersive bands in the same model. 

This peculiar behaviour is a consequence of the geometric origin of flat-band superfluidity. The total superfluid weight can be split in conventional and geometric contributions $D_{\rm s} = D_{\rm s,conv}+D_{\rm s,geom}$. The conventional contribution $D_{\rm s,conv}\propto J$ depends only on the derivatives of the dispersions $\varepsilon_{n\vc{k}}$ while the geometric one $D_{\rm s,geom}\propto \Delta_A$ includes derivatives of the Bloch functions $\ket{g_{n\vc{k}}}$~\cite{suppl}. Obviously the flat band does not contribute to the conventional term, while $D_{\rm s,geom} = \left.D_{\rm s,geom}\right|_{\rm f.b.}+\left.D_{\rm s,geom}\right|_{\rm o.b.}$ can be further split into a term originating purely from the flat band $\left.D_{\rm s,geom}\right|_{\rm f.b.} = \left.D_{\rm s}\right|_{\rm f.b.}$ and the remaining part $\left.D_{\rm s,geom}\right|_{\rm o.b.}$, which includes the geometric effect of the other bands. All three terms $D_{\rm s,conv},\,\left.D_{\rm s,geom}\right|_{\rm f.b.}$ and$\,\left.D_{\rm s,geom}\right|_{\rm o.b.}$ are invariant with respect to the gauge freedom consisting in the multiplication of the Bloch functions by an arbitrary $\vc{k}$-dependent phase factor and are thus well-defined. In our model the flat-band term $\left.D_{\rm s}\right|_{\rm f.b.}$ at half filling has the form
\begin{equation}
\label{flatband_cont}
\left.[D_{\rm s}]_{i,j}\right|_{\rm f.b.} = \frac{4}{\pi \hbar^2}\frac{\Delta_A^2}{U}\mathcal{M}_{ij}^{\rm R}|_{\rm f. b.} \approx \frac{U}{4\pi\hbar^2}\mathcal{M}_{ij}^{\rm R}|_{\rm f. b.}\,,
\end{equation}
where $\left.\mathcal{M}_{ij}^{\rm R}\right|_{\rm f.b.} = (2\pi)^{-1}\int_{\rm B.Z.}d^2\vc{k}\, \mathrm{Re}\left.\mathcal{B}_{ij}(\vc{k})\right|_{\rm f.b.}$ is the Brillouin-zone integral of the flat-band quantum metric $\mathrm{Re}\,\mathcal{B}_{ij}(\vc{k})|_{\rm f.b.}$. The quantum metric is defined as the real part of the quantum geometric tensor~\cite{Resta:2011,Provost:1980,Berry:1989}
\begin{equation}\label{eq:B}
\left.\mathcal{B}_{ij}(\vc{k})\right|_{\rm f.b.} = 2\langle \partial_{k_i} g_{0\vc{k}} |\big(1 - | g_{0\vc{k}} \rangle \langle g_{0\vc{k}}  |\big)| \partial_{k_j} g_{0\vc{k}} \rangle.
\end{equation}
It is worth mentioning that the same quantity $\mathcal{M}^{\rm R}$ appears in the theory of the polarization~\cite{Souza:2000,Resta:2011} and current~\cite{Neupert:2013} fluctuations in band insulators.
\begin{figure}[h!]
  \centering
    \includegraphics[width=0.25\textwidth]{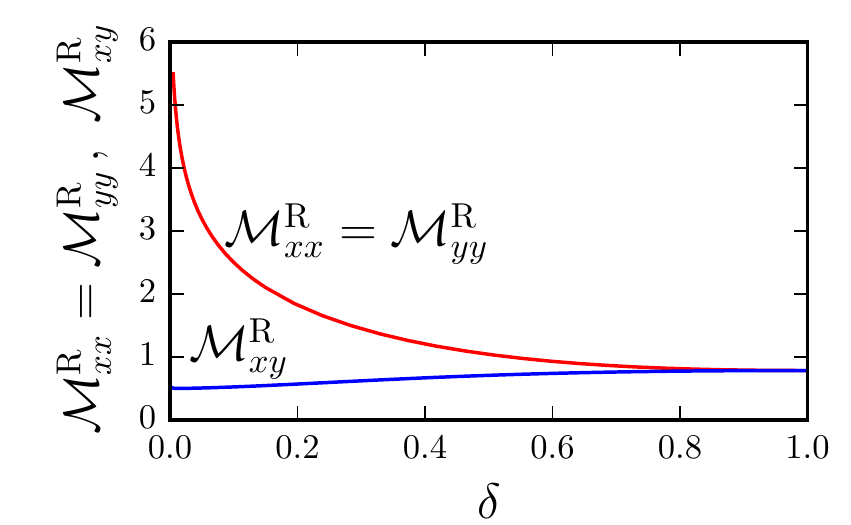}
   \caption{Brillouin-zone integral of the flat-band quantum metric $\left.\mathcal{M}^{\rm R}_{ij}\right|_{\rm f.b.}$ as a function of $\delta$. The diagonal components have a logarithmic singularity at $\delta=0$.}
   \label{fig:five}
\end{figure}
The strong dependence of $D_{\rm s}$ on $U$ for a partially filled flat band originates from the geometric term as shown in Figs.~\ref{fig:four}(a)-(b) where $D_{\rm s,conv}$ and $D_{\rm s,geom}$ are presented as a function of $U$ for half-filled flat band [$\nu = 1.5$, Fig.~\ref{fig:four}(a)] and half-filled upper band [$\nu = 2.5$, Fig.~\ref{fig:four}(b)]. For $\nu = 1.5$ the term $D_{\rm s,geom}$ dominates $D_{\rm s,conv}$, while for $\nu = 2.5$ $D_{\rm s,geom}$ is negligible at weak coupling.

In order to confirm the behavior of $D_{\rm s}$ observed in the mean-field calculations, we compute the Drude weight $D$ by using ED on periodic finite-size clusters of $12$, $18$, and $24$ sites~\cite{Dagotto:1992,suppl}. In the bulk limit $D$ is equivalent to $D_{\rm s}$ for gapped systems~\cite{Scalapino:1992,Scalapino:1993,Denteneer:1994}. Figs.~\ref{fig:four}(a)-(b) show that $D_s$ from BCS theory is in good agreement with ED results. In particular, at half filling ($\nu=1.5$), the sharp increase of $D$ for $0 \leq U \lesssim 4J$ becomes clearer with increasing cluster size. It is also peaked at $U\sim 4J$ and decreases when $U$ further increases, confirming the overall behavior of the mean-field $D_{\rm s}$. The drastic difference between $\nu=1.5$ and $2.5$ in the small $U$ limit is also confirmed by ED. The finite $D$ for $\nu=2.5$ at small coupling shows very weak dependence on $U$ for cluster size up to 24 sites.

In Figs.~\ref{fig:four}(c)-(d) we compare the conventional term $D_{\rm s,conv}$, the flat-band contribution $\left.D_{\rm s}\right|_{\rm f.b.}$ and the geometric contribution due to the other bands $\left.D_{\rm s,geom}\right|_{\rm o.b.}$ at half-filling $\nu = 1.5$ and for small $U \leq 0.2J$. Two values of $\delta$ are shown: $\delta = 5\cdot 10^{-3}$ [Fig.~\ref{fig:four}(c)] and $\delta = 0.1$ [Fig.~\ref{fig:four}(d)].
In both cases $D_{\rm s,conv}$ is negligible due to the vanishing density of states of the dispersive bands, while $\left.D_{\rm s}\right|_{\rm f.b.}$ gives the dominant contribution, linear in $U$. The term $\left.D_{\rm s,geom}\right|_{\rm o.b.}$ is negative and less relevant when the flat band becomes more isolated for larger $\delta$. When $U$ increases, the negative contribution of $\left.D_{\rm s,geom}\right|_{\rm o.b.}$ becomes more prominent and for very large $U$ it cancels the positive term $\left.D_{\rm s}\right|_{\rm f.b.}$ [see Fig.~\ref{fig:four}(a)].
This means that pairing has to occur in a subset of all bands for $D_{\rm s,geom}$ to manifest, and it explains the decreasing trend of $D_s$ in Figs.~\ref{fig:four}(a)-(b). As shown in Fig.~\ref{fig:five}, the invariant $\left.\mathcal{M}^{\rm R}_{ij}\right|_{\rm f.b.}$ diverges at $\delta = 0$, thus the slope of $D_{\rm s}$ as a function of $U$ is infinite at $U = 0$. However for any nonzero $U$ we have verified that this divergence is cured by $\left.D_{\rm s,geom}\right|_{\rm o.b.}$. Thus for $\delta = 0$ superfluidity has a truly multiband character. In the opposite limit $\delta \to 1$ one eigenvalue of $\left.\mathcal{M}^{\rm R}_{ij}\right|_{\rm f.b.}$ becomes zero and superfluidity is lost, consistently with the fact that the unit cells become decoupled [see Fig.~\ref{fig:one}(a)]. In contrast to mean-field theory, DMFT captures this behavior already at the level of the order parameter, as seen in Fig. \ref{fig:two}.

\noindent\textit{Discussion ---} The main result of this work is that topologically trivial flat bands are promising for high-$T_{\rm c}$ superconductivity, in the same way as topologically nontrivial ones. Indeed a flat band allows to optimize not only the BCS critical temperature~\cite{Noda:2015a}, but also the superfluid weight [see Figs.~\ref{fig:four}(a)-(b)]. The superfluid weight affects the critical temperature in two dimensions through the Berezinsky-Kosterlitz-Thouless (BKT) transition. We show that the superfluid weight has geometric origin, i.e. it is proportional to the quantum metric of the flat band [Eqs.~(\ref{flatband_cont})-(\ref{eq:B})]. The fingerprint of the geometric origin is the strong dependence of $D_{\rm s}$ on the coupling constant $U$, possibly observable in ultracold gases where interactions are tunable. 

Achieving the superfluid phase of an ultracold gas in an optical lattice is difficult, due to the still too high temperatures (specific entropies)  currently attainable~\cite{Esslinger:2010,Jordens:2010}. We find the BKT transition temperature in the Lieb lattice to be $k_{\rm B}T_{\rm c, BKT}=0.133 J$~\cite{suppl} at the optimal coupling $U \approx 4J$ [Fig.~\ref{fig:four}(a)]. It can be compared with the optimal N\'eel temperature for the 3D repulsive Fermi-Hubbard model $k_{\rm B}T_{\rm N\acute{e}el} = 0.333(7)J$~\cite{Fuchs:2011}, which is at the verge of experimental capabilities~\cite{Jordens:2010}. The critical temperatures are substantially higher in three dimensions where, in contrast to the BKT estimate in 2D, one can use the BCS one: $k_{\rm B}T_{\rm c, BCS} \approx 0.5-0.8 J$ for $U \approx 4 J$, and $\nu = 1.5$ in our model. The flat band optimizes the critical temperature, indeed $T_{\rm c, BKT}$ for the flat-band superfluid is twice as high compared to the dispersive bands in our model.

In the solid state context the geometric contribution to the superfluid weight is expected to be larger for superconductors with high-$T_{\rm c}$ and provides a possible explanation of the linear relation between superfluid weight and critical temperature in cuprates (Uemura relation~\cite{Uemura:1989,Marchetti:2015}) since $D_{{\rm s, geom}}\propto \Delta \propto T_{\rm c}$. We expect $D_{\textrm{s, geom}}$ to be significant in models with nontrivial Bloch functions also with the different pairing symmetries found in high-$T_{\rm c}$ superconductors, whose incorporation to our theory for the superfluid weight is an important topic of future research.


\begin{acknowledgments}
\noindent This work was supported by the Academy of Finland through
its  Centers  of  Excellence  Programme  (2012-2017)  and
under  Project  Nos.  263347,  251748, and  272490,  and
by  the  European  Research  Council  (ERC-2013-AdG-340748-CODE). This project has received funding from the European Union's Horizon 2020 research and innovation programme under the Marie Sklodowska-Curie grant agreement No 702281 (FLATOPS). We acknowledge useful discussions with Jildou Baarsma, Jami Kinnunen, Long Liang and Ari Harju. S. P. thanks Shunji Tsuchiya  for sharing his unpublished data, and both him and Wilhelm Zwerger for interesting discussions. T.I.V. is grateful for the support from the Vilho, Yrj\"{o} and Kalle V\"{a}is\"{a}l\"{a} Foundation. D.H.K. acknowledges support from the National Research Foundation of Korea through its Basic Science Research Program (NRF-2014R1A1A1002682, NRF-2015K2A7A1035792). Computing resources were provided by CSC -- the Finnish IT Centre for Science and the Triton cluster at Aalto University.
\end{acknowledgments}

\bibliographystyle{apsrev4-1}

\clearpage
\onecolumngrid
\appendix

\section{Appendix A: Multiband BCS approach and superfluid weight}

\subsection{1. Bogoliubov de-Gennes Hamiltonian}\label{section:bdghamiltonian}

In order to calculate the superfluid weight we use the multiband BCS theory developed in Ref.~\cite{Sebastiano:2015}, which is a mean-field approach, with the difference that we take into account the Hartree potentials $n_{\alpha} = \langle \hat{n}_{\vc{i}\alpha\sigma}\rangle$ ($\alpha \in \{A,B,C \}$) of Eq.~(2) in the main text, as explained below. Furthermore, the Lieb lattice geometry and band structure are used. The general procedure is repeated here for completeness. The starting point is the Bogoliubov-de Gennes Hamiltonian $\bar{H}_{\vc{k}}(\vc{q})$ in the presence of a pairing amplitude $\Delta(\vc{r}) = \Delta e^{2i\vc{q}\cdot \vc{r}}$ with finite phase winding given by the wavevector $\vc{q}$. This nonuniform phase describes a state with a finite supercurrent. The corresponding Bogoliubov-de Gennes Hamiltonian is
\begin{equation}\label{eq:BdG}
\bar{H}_{\vc{k}}(\vc{q}) = \begin{pmatrix}
\varepsilon_{\vc{k}-\vc{q}}-\mu \bm{1}& \mathcal{G}_{\vc{k}-\vc{q}}^\dagger\Delta\mathcal{G}_{\vc{k}+\vc{q}} \\
\mathcal{G}_{\vc{k}+\vc{q}}^\dagger\Delta\mathcal{G}_{\vc{k}-\vc{q}}  & -\left(\varepsilon_{\vc{k}+\vc{q}}-\mu \bm{1}\right)
\end{pmatrix}\,, \qquad \text{with} \quad \Delta = 
\begin{pmatrix} \Delta_A & 0 & 0 \\
0 & \Delta_B  & 0 \\
0 & 0 & \Delta_A
\end{pmatrix}\,.
\end{equation}
Note that the Bogoliubov-de Gennes Hamiltonian is a $2\times 2$ block matrix, where the diagonal matrices of the band dispersions $\varepsilon_{\vc{k}} = \mathrm{diag}(\varepsilon_{n\vc{k}})$ and the unitary matrix of the Bloch functions $g_{n\vc{k}}(\alpha) = [\mathcal{G}_{\vc{k}}]_{\alpha,n}$ are obtained by diagonalizing the kinetic single-particle Hamiltonian $H_{\vc{k}} = \mathcal{G}_{\vc{k}}\varepsilon_{\vc{k}}\mathcal{G}_{\vc{k}}^\dagger$. The notation $[M]_{a,b}$ for the matrix elements of a matrix $M$ is used throughout. The kinetic single-particle Hamiltonian reads 
\begin{align}
\label{fourierhamiltonian}
H_{\vc{k}}  = 2J 
\begin{pmatrix}
\gamma_A/2 & a_{\vc{k}} & 0 \\
a_{\vc{k}}^* & \gamma_B/2 &b_{\vc{k}}  \\
0  & b_{\vc{k}}^* & \gamma_A/2
\end{pmatrix}\,.
\end{align}
This is the same as Eq.~(1) in the main text with the only difference that the Hartree potential $\gamma_\alpha = -U/J(n_\alpha-1/2)$ has been included. Then the energy dispersions and the Bloch functions in the presence of a finite Hartree term are
\begin{gather}
\varepsilon_{\vc{k}} = 
\begin{pmatrix}
\varepsilon_{+,\vc{k}} & & \\
& J\gamma_A& \\
& & \varepsilon_{-,\vc{k}} 
\end{pmatrix}\\
\varepsilon_{\pm, \vc{k}} =  J(\gamma_A + \gamma_B) \pm 2J\sqrt{(\gamma_A+\gamma_B)^2 + |a_{\vc{k}}|^2 + |b_{\vc{k}}|^2}\\
\mathcal{G}_{\vc{k}} = 
\begin{pmatrix}
\frac{a_{\vc{k}}}{\sqrt{2\chi_{\vc{k}} + 2(\gamma_A-\gamma_B)\sqrt{\chi_{\vc{k}}}}} & -\frac{b_{\vc{k}}}{\sqrt{|a_{\vc{k}}|^2+|b_{\vc{k}}|^2}} & \frac{a_{\vc{k}}}{\sqrt{2\chi_{\vc{k}} - 2(\gamma_A-\gamma_B)\sqrt{\chi_{\vc{k}}}}} \\
\frac{-\gamma_A+\gamma_B -\sqrt{\chi_{\vc{k}}}}{\sqrt{2\chi_{\vc{k}} + 2(\gamma_A-\gamma_B)\sqrt{\chi_{\vc{k}}}}}& 0 &\frac{-\gamma_A+\gamma_B +\sqrt{\chi_{\vc{k}}}}{\sqrt{2\chi_{\vc{k}} - 2(\gamma_A-\gamma_B)\sqrt{\chi_{\vc{k}}}}} \\
\frac{b_{\vc{k}}^*}{\sqrt{2\chi_{\vc{k}} + 2(\gamma_A-\gamma_B)\sqrt{\chi_{\vc{k}}}}} & \frac{a^*_{\vc{k}}}{\sqrt{|a_{\vc{k}}|^2+|b_{\vc{k}}|^2}} & \frac{b_{\vc{k}}^*}{\sqrt{2\chi_{\vc{k}} - 2(\gamma_A-\gamma_B)\sqrt{\chi_{\vc{k}}}}}
\end{pmatrix}\,, \label{eq:G_def}
\end{gather}
where $\chi_{\vc{k}} = (\gamma_A-\gamma_B)^2 +|a_{\vc{k}}|^2 + |b_{\vc{k}}|^2$. Note that the Hartree potential has the effect of shifting the flat band energy, but the flat-band Bloch functions (the middle column of Eq.~(\ref{eq:G_def})) are unaffected.

In general the pairing potentials $\Delta_\alpha$ and the Hartree potentials $n_\alpha$ have to be found self-consistently for any value of $\vc{q}$. According to Ref.~\cite{Sebastiano:2015} it is necessary to find the self-consistent solution only for $\vc{q} = 0$ in order to calculate the superfluid density. The diagonalization of Eq.~\eqref{eq:BdG} for $\vc{q}=0$ provides the quasiparticle energies ($E_{n\vc{k}}$) and wavefunctions ($\mathcal{W}_{\vc{k}}$)
\begin{equation}
\bar{H}_{\vc{k}}(\vc{q}=0) = \mathcal{W}_{\vc{k}}(\vc{q}=0) E_{\vc{k}}(\vc{q}=0)\mathcal{W}_{\vc{k}}^\dag(\vc{q}=0)\,,
\end{equation}
with 
\begin{equation}
\label{quasienergies}
 E_{\vc{k}}(\vc{q}=0)= \begin{pmatrix}
E^>_{\vc{k}} & 0\\
 0 & -E^>_{\vc{k}} \\
\end{pmatrix}  = \begin{pmatrix}
\textrm{diag}(E_{n\vc{k}}) & 0\\
 0 & -\textrm{diag}(E_{n\vc{k}}) \\
\end{pmatrix}
\end{equation}
and 
\begin{equation}
 \mathcal{W}_{\vc{k}}(\vc{q}=0) = \begin{pmatrix}
\mathcal{U}_{\vc{k}} & -\mathcal{V}_{\vc{k}}\\
\mathcal{V}_{\vc{k}} & \mathcal{U}_{\vc{k}} \\
\end{pmatrix}.
\end{equation}
We use the notation $\mathcal{U}_{\vc{k}},\,\mathcal{V}_{\vc{k}}$ for the blocks of $ \mathcal{W}_{\vc{k}}(\vc{q}=0)$ as a reminder that these are the generalization for a multiband Bogoliubov-de Gennes Hamiltonian of the usual BCS coherence factors $u_{\vc{k}},\,v_{\vc{k}}$~\cite{Grosso_Book}.
In Eq.~\eqref{quasienergies} the diagonal matrix $E^>_{\vc{k}}$ contains the positive quasiparticle energies $E_{n\vc{k}}\geq 0$. 

\subsection{2. General expression for the superfluid weight tensor in a multiband system}
For convenience we define the following quantities:
\begin{gather}
\mathcal{D}_{\vc{k}}(\vc{q}) = - \mathcal{G}_{\vc{k}-\vc{q}}^\dagger\Delta\mathcal{G}_{\vc{k}+\vc{q}} = \mathcal{D}^{\dagger}_{\vc{k}}(-\vc{q}) \label{eq:D_def}\\
\begin{split}
&N_{\vc{k},i} = \mathcal{W}_{\vc{k}}^\dagger(\vc{q}=0)\, \partial_{q_i}\bar{H}_{\vc{k}}(\vc{q}=0)\,
\mathcal{W}_{\vc{k}}(\vc{q}=0) = 
\begin{pmatrix}
A_{\vc{k},i} & B_{\vc{k},i} \\ -B_{\vc{k},i} & A_{\vc{k},i}
\end{pmatrix} \\ 
&\text{with}\;\;
\begin{cases}
A_{\vc{k},i} = \mathcal{U}_{\vc{k}}^\dagger\partial_{k_i}\varepsilon_{\vc{k}} \mathcal{U}_{\vc{k}} + \mathcal{V}_{\vc{k}}^\dagger\partial_{k_i}\varepsilon_{\vc{k}} \mathcal{V}_{\vc{k}} +
\mathcal{U}_{\vc{k}}^\dagger \partial_{q_i}\mathcal{D}_{\vc{k}}(\vc{q}=0) \mathcal{V}_{\vc{k}} - \mathcal{V}_{\vc{k}}^\dagger \partial_{q_i}\mathcal{D}_{\vc{k}}(\vc{q}=0) \mathcal{U}_{\vc{k}} 
\\
B_{\vc{k},i} = \mathcal{U}_{\vc{k}}^\dagger \partial_{q_i}\mathcal{D}_{\vc{k}}(\vc{q}=0) \mathcal{U}_{\vc{k}} + \mathcal{V}_{\vc{k}}^\dagger \partial_{q_i}\mathcal{D}_{\vc{k}}(\vc{q}=0) \mathcal{V}_{\vc{k}} 
+ \mathcal{V}_{\vc{k}}^\dagger\partial_{k_i}\varepsilon_{\vc{k}} \mathcal{U}_{\vc{k}} - \mathcal{U}_{\vc{k}}^\dagger\partial_{k_i}\varepsilon_{\vc{k}} \mathcal{V}_{\vc{k}}
\end{cases}
\end{split}
\label{eq:aux_def1}
\\
[T_{\vc{k}}]_{a,b} = 
\begin{cases}
\left[\dfrac{\beta}{2\cosh^2(\beta E_{\vc{k}}/2)}\right]_{a,a} & \text{for}\quad a = b\,, \\[1.5em]
\dfrac{[\tanh(\beta E_{\vc{k}}/2)]_{a,a}-[\tanh(\beta E_{\vc{k}}/2)]_{b,b}}{[E_{\vc{k}}]_{a,a}-[E_{\vc{k}}]_{b,b}} & \text{for}\quad a \neq b\,.
\end{cases}\label{eq:aux_def2}
\end{gather}
Using these definitions it is possible to derive the following result for the superfluid weight tensor $[D_s]_{i,j}$
\begin{gather}\label{eq:superfluid_weight_final_1}
[D_{\rm s}]_{i,j} = \left.\frac{1}{V\hbar^2}\frac{\partial^2\Omega}{\partial q_i\partial q_j}\right|_{_{\mu,\Delta,\vc{q}=0}} = [D_{\text{s,conv}}]_{i,j} 
+ [D_{\text{s,geom}}]_{i,j}\,, \qquad \text{with}\\
[D_{\text{s,conv}}]_{i,j} =  \frac{2}{V\hbar^2}\sum_{\vc{k}}\mathrm{Tr}\left[\left(\mathcal{V}_{\vc{k}}\frac{1}{e^{-\beta E_{\vc{k}}^>}+1}\mathcal{V}_{\vc{k}}^\dagger +
\mathcal{U}_{\vc{k}}\frac{1}{e^{\beta E_{\vc{k}}^>}+1}\mathcal{U}_{\vc{k}}^\dagger \right)\partial_{k_i}\partial_{k_j}\varepsilon_{\vc{k}}\right]\,, \label{eq:superfluid_weight_final_2}\\ 
[D_{\text{s,geom}}]_{i,j} = \frac{1}{V\hbar^2}\Big\{ 2\sum_{\vc{k}}\mathrm{Tr}\left[\left(\mathcal{U}_{\vc{k}}\mathcal{V}_{\vc{k}}^\dagger -\mathcal{U}_{\vc{k}}\frac{1}{e^{\beta E^>_{\vc{k}}}+1}\mathcal{V}^\dagger_{\vc{k}}-\mathcal{V}_{\vc{k}}\frac{1}{e^{\beta E^>_{\vc{k}}}+1}\mathcal{U}_{\vc{k}}^\dagger\right)\partial_{q_i}\partial_{q_j}\mathcal{D}_{\vc{k}}
(\vc{q}=0)\right] \nonumber \\
-\frac{1}{2}\sum_{\vc{k}} \sum_{a,b} [T_{\vc{k}}]_{a,b}[N_{\vc{k},i}]_{a,b}[N_{\vc{k},j}]_{b,a}\Big\}\,.\label{eq:superfluid_weight_final_3}
\end{gather}
The superfluid weight tensor is defined as the derivatives with respect to $\vc{q}$ of the grand potential $\Omega(\mu,T,\Delta,\vc{q})$ and $V$ is the system volume (area in 2D). We use here a different notation than in Ref.~\cite{Sebastiano:2015}: the conventional contribution to the superfluid weight $D_{\text{s,conv}}$ is called $D_{\text{s},1}$ in Ref.~\cite{Sebastiano:2015}, while the geometric one $D_{\text{s,geom}}$ corresponds to $D_{{\rm s},2} + D_{{\rm s},3}$ in the same reference. The conventional contribution is distinguished by the fact that only the derivatives of the band dispersion enter in Eq.~\eqref{eq:superfluid_weight_final_2}, while in the geometric one also the derivatives of the Bloch functions appear through the quantities $\partial_{q_i}\mathcal{D}_{\vc{k}}(\vc{q}=0)$, $\partial_{q_i}\partial_{q_j}\mathcal{D}_{\vc{k}}(\vc{q}=0)$, where $\mathcal{D}_{\vc{k}}(\vc{q})$ is defined in Eq.~\eqref{eq:D_def}. Moreover, the only energy scale of the conventional contribution is the hopping energy $J$, which is the scale of the band dispersion $\varepsilon_{\vc{k}}$. On the other hand, $D_{\text{s,geom}}$ depends also on the energy gaps $\Delta_\alpha$ again through $\mathcal{D}_{\vc{k}}(\vc{q})$.

\subsection{3. Changing the filling within the flat band}

As the formulas for the superfluid weight are derived in the grand canonical ensemble, the chemical potential $\mu$ is fixed rather than the total filling $\nu = \sum_\alpha n_\alpha$. In case of dispersive bands we can scan the filling by simply changing the chemical potential. However, in case of the flat band the same chemical potential $\mu=0$ corresponds to an arbitrary partial filling of the flat band, namely the filling $\nu(\mu)$ as a function of $\mu$ is discontinuous at $\mu = 0$. In order to obtain the superfluid weight as a function of filling presented in Fig.~3 of the main text, we exploit the fact that once a self-consistent solution for $\mu = 0$ is found, which corresponds to a partially filled flat band, it is possible to obtain another self-consistent solution by an arbitrary rotation of the following three dimensional vectors~\cite{montecarlo} 
\begin{align}
\vc{S}_\alpha = \Big(
\text{Re\Big[}\frac{\Delta_\alpha}{-U}\Big],
\text{Im}\Big[\frac{\Delta_\alpha}{-U}\Big],
n_\alpha - \frac{1}{2} \Big)\,.
\end{align}
The rotation is the same for all sublattices labelled by $\alpha$. This is a fundamental symmetry of any bipartite lattice and it can be better appreciated by performing the particle-hole transformation introduced by Emery that maps the attractive Hubbard model into the repulsive one~\cite{Emery:1976}. In the case of the repulsive Hubbard model, this symmetry corresponds to rotations of the magnetization vector.
Note that this symmetry holds only if the pairing potentials $\Delta_\alpha$ and Hartree potentials $n_\alpha$ are treated on an equal footing. This is the reason to introduce the Hartree potential.
By employing this symmetry, we are able to obtain the superfluid weight for any filling of the flat band. 

\subsection{4. Analytical results for half-filled flat band}

In general we adopt a fully numerical approach to solve the self-consistent equations given in Ref.~\cite{Sebastiano:2015} and evaluate the superfluid weight from Eqs.~\eqref{eq:superfluid_weight_final_1}-\eqref{eq:superfluid_weight_final_3}.
However, it turns out that an analytical solution can be found when the flat band is half-filled. The theorem of Ref.~\cite{Lieb:1993} guarantees that the Hartree potential vanishes precisely at half-filling ($n_{\alpha} - 1/2 = 0 = \gamma_{\alpha}$). The quasiparticle energies, the eigenvalues of \eqref{eq:BdG}, take a very simple form at half filling
\begin{equation}
E_{\pm,\vc{k}} = \sqrt{\epsilon_{\vc{k}}^2 + \Delta_{\rm s}^2} \pm |\Delta_{\rm d}| \geq 0\,,\quad E_{0,\vc{k}} = \Delta_A\,,\quad E_{\vc{k}}^> = \mathrm{diag}(E_{+,\vc{k}},E_{0,\vc{k}},E_{-,\vc{k}})\,,
\end{equation}
where $\Delta_{\rm{d}} = (\Delta_A-\Delta_B)/2$, $\Delta_{\rm{s}} = (\Delta_A+\Delta_B)/2$ and $\epsilon_{\vc{k}} = 2J\sqrt{|a_{\vc{k}}|^2+|b_{\vc{k}}|^2}$.
Correspondingly, the unitary matrix  $\mathcal{W}_{\vc{k}}$ that diagonalizes the BdG Hamiltonian reads
\begin{equation}\label{eq:W_half}
\mathcal{W}_{\vc{k}} =
\begin{pmatrix}
\mathcal{U}_{\vc{k}} & -\mathcal{V}_{\vc{k}} \\
\mathcal{V}_{\vc{k}} & \mathcal{U}_{\vc{k}}
\end{pmatrix}\,,\qquad 
\mathcal{U}_{\vc{k}} = \frac{1}{\sqrt{2}}
\begin{pmatrix}
\cos \frac{\phi_{\vc{k}}}{2} & 0 &-\cos\frac{\phi_{\vc{k}}}{2} \\
0 & 1 & 0 \\
\sin \frac{\phi_{\vc{k}}}{2} & 0 & \sin \frac{\phi_{\vc{k}}}{2}
\end{pmatrix}\,, \qquad 
\mathcal{V}_{\vc{k}} = \frac{1}{\sqrt{2}}
\begin{pmatrix}
\sin \frac{\phi_{\vc{k}}}{2} & 0 & -\sin\frac{\phi_{\vc{k}}}{2} \\
0 & 1 & 0 \\
\cos \frac{\phi_{\vc{k}}}{2} & 0 & \cos \frac{\phi_{\vc{k}}}{2}
\end{pmatrix}\,.
\end{equation}
Here the coefficients of $\mathcal{U}_{\vc{k}}$ and $\mathcal{V}_{\vc{k}}$ take precisely the form of BCS coherence factors
\begin{equation}\label{eq:sin_cos_half}
\cos \frac{\phi_{\vc{k}}}{2} = \frac{1}{\sqrt{2}}\sqrt{1+\frac{\epsilon_{\vc{k}}}{\sqrt{\epsilon_{\vc{k}}^2 + \Delta_{\rm s}^2}}}\,,\qquad \sin \frac{\phi_{\vc{k}}}{2} = \frac{1}{\sqrt{2}}\sqrt{1-\frac{\epsilon_{\vc{k}}}{\sqrt{\epsilon_{\vc{k}}^2 + \Delta_{\rm s}^2}}}\,.
\end{equation}
Away from half-filling the block stucture of $\mathcal{U}_{\vc{k}}$ and $\mathcal{V}_{\vc{k}}$ survives, namely the flat band, which corresponds to the middle $1\times 1$ block in Eq.~\eqref{eq:W_half}, is decoupled from the other bands for any filling and the corresponding $2\times 2$ Bogoliubov-de Gennes Hamiltonian can be trivially solved. This is a peculiar feature of our model, which implies that there is a flat band of quasiparticle excitations. 


Given the above results for the eigenvectors $\mathcal{W}_{\vc{k}}$ and the eigenvalues $E_{n\vc{k}}$, the only ingredient needed for the evaluation of the superfluid weight are the derivatives of the matrix $\mathcal{D}_{\vc{k}}(\vc{q}) = -\mathcal{G}_{\vc{k-q}}^\dagger\Delta \mathcal{G}_{\vc{k+q}}$.
Let us introduce a two-component complex spinor $\ket{s_{\vc{k}}}$ and its partner obtained by time-reversal symmetry $\ket{\bar{s}_{\vc{k}}}$
\begin{equation}\label{eq:spinor}
\ket{s_{\vc{k}}} 
= \frac{1}{\sqrt{|a_{\vc{k}}|^2+|b_{\vc{k}}|^2}}
\begin{pmatrix}
a_{\vc{k}} \\ b_{\vc{k}}^* 
\end{pmatrix}\,,\qquad 
\ket{\bar{s}_{\vc{k}}} = \mathcal{T}\ket{s_{\vc{k}}} 
= i\sigma_y\mathcal{C}\ket{s_{\vc{k}}}
= \frac{1}{\sqrt{|a_{\vc{k}}|^2+|b_{\vc{k}}|^2}}
\begin{pmatrix}
b_{\vc{k}} \\ -a_{\vc{k}}^* 
\end{pmatrix}\,.
\end{equation}
Here $\mathcal{T} = i\sigma_y\mathcal{C}$ is the time reversal operator and $\mathcal{C}$ is the complex conjugate operator. It follows from the definitions that $\braket{s_{\vc{k}}}{\bar{s}_{\vc{k}}} = 0$. The spinor $\ket{s_{\vc{k}}}$ is a purely formal construction and does not have any direct physical meaning. Using these definitions the matrix $\mathcal{G}_{\vc{k}_1}^\dagger\Delta \mathcal{G}_{\vc{k}_2}$ can be represented as
\begin{equation}\label{eq:GDG}
\mathcal{G}_{\vc{k}_1}^\dagger\Delta \mathcal{G}_{\vc{k}_2} = 
\frac{\Delta_A}{2}\begin{pmatrix}
\braket{s_{\vc{k}_1}}{s_{\vc{k}_2}} & \sqrt{2}\braket{s_{\vc{k}_1}}{\bar{s}_{\vc{k}_2}} & \braket{s_{\vc{k}_1}}{s_{\vc{k}_2}} \\
\sqrt{2}\braket{\bar{s}_{\vc{k}_1}}{s_{\vc{k}_2}} & 
2\braket{\bar{s}_{\vc{k}_1}}{\bar{s}_{\vc{k}_2}} & \sqrt{2}\braket{\bar{s}_{\vc{k}_1}}{s_{\vc{k}_2}}\\
\braket{s_{\vc{k}_1}}{s_{\vc{k}_2}} & 
\sqrt{2}\braket{s_{\vc{k}_1}}{\bar{s}_{\vc{k}_2}} & 
\braket{s_{\vc{k}_1}}{s_{\vc{k}_2}}
\end{pmatrix}+
\frac{\Delta_B}{2}\begin{pmatrix}
1 & 0 & -1 \\
0 & 0 & 0 \\
-1 & 0 & 1
\end{pmatrix}
\,.
\end{equation}
Note that the matrix $\mathcal{G}_{\vc{k}_1}^\dagger\Delta \mathcal{G}_{\vc{k}_2}$ is the sum of two terms proportional to the order parameters $\Delta_A$ and $\Delta_B$, respectively. Only the term proportional to $\Delta_A$ depends on the wavevectors $\vc{k}_{1,2}$. Therefore the derivatives of the matrix $\mathcal{D}_{\vc{k}}(\vc{q}) = -\mathcal{G}_{\vc{k-q}}^\dagger\Delta \mathcal{G}_{\vc{k+q}}$ are equal to the derivatives of $\mathcal{D}'_{\vc{k}}(\vc{q}) = -\Delta_A\mathcal{G}_{\vc{k-q}}^\dagger \mathcal{G}_{\vc{k+q}}$, i.e. one can set $\Delta_B = \Delta_A$ for the purpose of calculating derivatives. As shown in Ref.~\cite{Sebastiano:2015}, this provides a number of simplifications. As a consequence only the energy scale $\Delta_A =\Delta_C$ enters in the geometric contribution to the superfluid weight, but not $\Delta_B$.
Another advantage of Eq.~(\ref{eq:GDG}) is that the calculation of the derivatives of the six independent matrix elements of a $3\times 3$ hermitian matrix is reduced to the calculation of the derivatives of only two quantities, namely $\braket{s_{\vc{k}_1}}{s_{\vc{k}_2}} = \braket{\bar{s}_{\vc{k}_1}}{\bar{s}_{\vc{k}_2}}^*$ and $\braket{s_{\vc{k}_1}}{\bar{s}_{\vc{k}_2}} =  - \braket{\bar{s}_{\vc{k}_1}}{s_{\vc{k}_2}}^*$. The quantum geometric tensor of the flat band reads in the spinor notation
\begin{equation}
\left.\mathcal{B}_{ij}(\vc{k})\right|_{\rm f. b.} = 2\braket{\partial_{k_i}\bar{s}_{\vc{k}}}{s_{\vc{k}}}\braket{s_{\vc{k}}}{\partial_{k_j}\bar{s}_{\vc{k}}} = 2\braket{\partial_{k_j}s_{\vc{k}}}{\bar{s}_{\vc{k}}}\braket{\bar{s}_{\vc{k}}}{\partial_{k_i}s_{\vc{k}}} =2\langle \partial_{k_i} g_{0\vc{k}} |\big(1 - | g_{0\vc{k}} \rangle \langle g_{0\vc{k}}  |\big)| \partial_{k_j} g_{0\vc{k}} \rangle\,.
\end{equation}
The real part of the quantum geometric tensor $\mathcal{B}_{ij}(\vc{k})$ is called the quantum metric.

\subsection{5. Gap equations at half filling}
Using Eqs.~(\ref{eq:W_half})-(\ref{eq:sin_cos_half}) and the general results of Ref.~\cite{Sebastiano:2015} one obtains the gap equations
\begin{gather}
\Delta_A = \Delta_C = \frac{U}{4N_{\rm c}}\sum_{\vc{k}}\left[t_{+,\vc{k}}\sin\phi_{\vc{k}}+t_{-,\vc{k}}\right]+{\color{BrickRed}\frac{U}{4}\tanh\frac{\beta\Delta_A}{2}}\,,\label{eq:gap_1_T}\\
\Delta_B = \frac{U}{2N_{\rm c}}\sum_{\vc{k}}\left[t_{+,\vc{k}}\sin\phi_{\vc{k}}-t_{-,\vc{k}}\right]\label{eq:gap_2_T}\,,\\
\text{with}\quad t_{\pm,\vc{k}} = \frac{1}{2}\left(\tanh\frac{\beta E_{+,\vc{k}}}{2}\pm\tanh\frac{\beta E_{-,\vc{k}}}{2}\right)\,.
\end{gather}
Here $N_{\rm c}$ is the number of unit cells in the lattice. In the zero temperature limit ($t_{+,\vc{k}}\to 1,\,\tanh(\beta\Delta_A/2) \to 1\,,t_{-,\vc{k}}\to 0$) the gap equations read
\begin{gather}
\Delta_A = \Delta_C = \frac{U}{4N_{\rm c}}\sum_{\vc{k}}
\frac{\Delta_{\rm s}}{\sqrt{\epsilon_{\vc{k}}^2+\Delta_{\rm s}^2}}+{\color{BrickRed}\frac{U}{4}}\,,\label{eq:gap_1}\\
\Delta_B = \frac{U}{2N_{\rm c}}\sum_{\vc{k}}\frac{\Delta_{\rm s}}{\sqrt{\epsilon_{\vc{k}}^2+\Delta_{\rm s}^2}}\,.\label{eq:gap_2}
\end{gather}
The gap equations for the two order parameters $\Delta_A$ and $\Delta_B$ are coupled since $\Delta_{\rm s} = (\Delta_A+\Delta_B)/2$. The flat band provides $\vc{k}$-independent terms in the gap equations for the order parameter $\Delta_A= \Delta_C$, namely the term $\frac{U}{4}\tanh\frac{\beta\Delta_A}{2}$ in Eq.~(\ref{eq:gap_1_T}) and $\frac{U}{4}$ in Eq.~(\ref{eq:gap_1}) (highlighted in {\color{BrickRed}red}). It makes sense that the the flat band enters only in the equations for the order parameter $\Delta_A$, but not $\Delta_B$, since the flat band is composed of states that are localized in the $A,C$ sublattices~\cite{montecarlo}.
From the zero temperature gap equations the asymptotic behaviour of the order parameters for small $U$ is derived
\begin{equation}
\Delta_A \approx \frac{n_{\phi}U}{2}\left(1+\frac{U}{8J}I(\delta)\right)\quad {\rm with}\,\,n_{\phi}^{-1}=2\,,\qquad \Delta_B \approx \Delta_A\frac{U}{4J}I(\delta) \approx \frac{n_\phi U^2}{8J}I(\delta)\,. \label{eq:Delta_small_U}
\end{equation}
The constant $I(\delta)$ is defined by
\begin{equation}
I(\delta) = \frac{J}{N_{\rm c}}\sum_{\vc{k}}\frac{1}{\epsilon_{\vc{k}}} = \int_{0}^{2\pi}dx\int_{0}^{2\pi}dy\,\frac{1}{2\sqrt{1+\delta^2+\frac{1-\delta^2}{2}(\cos x+\cos y)}}\,.
\end{equation}
For the value $\delta = 10^{-3}$ used in most of the calculations one obtains $I(\delta) \approx 0.64\,$. The leading order result for $\Delta_A = n_\phi U/2$ agrees with the general result in the isolated flat-band case~\cite{Sebastiano:2015}, where $n_{\phi}^{-1}= 2$ is the number of orbitals (sublattices) on which the flat-band states have nonvanishing amplitude. 

\subsection{6. Superfluid weight at half filling}\label{sec:sfw}

Using Eqs.~(\ref{eq:W_half})-(\ref{eq:sin_cos_half}) and after a straightforward but tedious calculation, one can derive the following expression for the superfluid weight as a summation (integral) of a function of $\vc{k}$ over the whole Brillouin zone ($A = N_{\rm c}a^2$ is the system area, $a$ the lattice constant)
\begin{gather}\label{eq:superfluid_final}
\begin{split}
[D_{\rm s}]_{i,j} = \frac{1}{A\hbar^2}\sum_{\vc{k}}&\bigg[-2t_{+,\vc{k}}\cos \phi_{\vc{k}}\partial_{k_i}\partial_{k_j}\epsilon_{\vc{k}}
-\frac{4t_{-,\vc{k}}}{E_{+,\vc{k}}-E_{-,\vc{k}}}\partial_{k_i}\epsilon_{\vc{k}}\partial_{k_j}\epsilon_{\vc{k}}\\
&+ 2{\Delta_A}\left({\color{BrickRed}\tanh\frac{\beta \Delta_A}{2}} + t_{+,\vc{k}}\sin\phi_{\vc{k}} + t_{-,\vc{k}}\right)\left(\braket{\partial_{k_i}s_{\vc{k}}}{\partial_{k_j}s_{\vc{k}}}+\braket{\partial_{k_j}s_{\vc{k}}}{\partial_{k_i}s_{\vc{k}}}\right)\\
&-\Delta_A^2 \braket{\partial_{k_i}s_{\vc{k}}}{s_{\vc{k}}}\braket{s_{\vc{k}}}{\partial_{k_j}s_{\vc{k}}}f(\vc{k})-\Delta_A^2\left(\braket{\partial_{k_i}s_{\vc{k}}}{\bar{s}_{\vc{k}}}\braket{\bar{s}_{\vc{k}}}{\partial_{k_j}s_{\vc{k}}}+(i\leftrightarrow j)\right)g(\vc{k})\bigg]\,.
 \end{split}
\end{gather}
where the functions $f(\vc{k})$ and $g(\vc{k})$ are defined as
\begin{gather}\label{eq:ff}
f(\vc{k}) = (1+\sin\phi_{\vc{k}})^2\frac{\tanh(\beta E_{+,\vc{k}}/2)}{E_{+,\vc{k}}}+(1-\sin\phi_{\vc{k}})^2\frac{\tanh(\beta E_{-,\vc{k}}/2)}{E_{-,\vc{k}}} + {\color{BrickRed}4\frac{\tanh(\beta\Delta_A/2)}{\Delta_A}} + 4\cos^2\phi_{\vc{k}}\frac{t_{-,\vc{k}}}{E_{+,\vc{k}}-E_{-,\vc{k}}}\,,\\
g(\vc{k}) = 2(1-\sin\phi_{\vc{k}})\frac{\tanh(\beta E_{-,\vc{k}}/2)-\tanh(\beta \Delta_A/2)}{E_{-,\vc{k}}-\Delta_A}
+ 2(1+\sin\phi_{\vc{k}})\frac{\tanh(\beta E_{+,\vc{k}}/2)+\tanh(\beta \Delta_A/2)}{E_{+,\vc{k}}+\Delta_A}\,.\label{eq:gg}
\end{gather}
One can distinguish two gauge-invariant superfluid weight contributions. The conventional contribution $D_{\text{s,conv}}$ is the one given by the first two terms in square brackets in Eq.~(\ref{eq:superfluid_final}) (first line).
At half-filling this contribution is highly suppressed due to the vanishing density of states of the dispersive bands as it can be seen in Figs. 4(c)-(d) in the main text.
The terms where the spinors $\ket{s_{\vc{k}}}$,$\ket{\bar{s}_{\vc{k}}}$ and their derivatives appear represent the geometric contribution $D_{\text{s,geom}}$. 
 
From Eqs.~(\ref{eq:superfluid_final})-(\ref{eq:gg}) it is possible to single out the flat-band contribution $D_{\rm s,geom}|_{\rm f.b.}=D_{\rm s}|_{\rm f.b.}$ to the superfluid weight (highlighted in {\color{BrickRed}red}) from the geometric contribution associated to the other bands $D_{\rm s,geom}|_{\rm o.b.}$. Formally, one considers the  isolated flat-band limit $0 < \frac{U}{J} \ll \delta < 1$ which means that pairing occurs in the flat band only. In this limit one can set $\sin\phi_{\vc{k}} = t_{-,\vc{k}} = 0$ and all terms of order $\Delta_A/E_{\pm,\vc{k}} \approx U/({J\delta})$ are discarded.
Then one obtains
\begin{equation}\label{eq:sfw_flat}
\begin{split}
[D_{{\rm s}}]_{i,j} &= [\left.D_{\text{s,geom}}\right|_{\rm f.b.} ]_{i,j}= \frac{\Delta_A}{\pi\hbar^2}\tanh\frac{\beta \Delta_A}{2}\frac{1}{2\pi}\int_{\rm B.Z.} d^2\vc{k}\,\mathrm{Re}\left.\mathcal{B}_{ij}(\vc{k})\right|_{\rm f. b.}
 \\ &= \frac{\Delta_A}{\pi\hbar^2}\tanh\frac{\beta \Delta_A}{2} \mathcal{M}_{ij}^{\rm R}|_{\rm f. b.} = \frac{2}{\pi \hbar^2}\frac{\Delta_A^2}{Un_{\phi}}\mathcal{M}_{ij}^{\rm R}|_{\rm f. b.}\,.
\end{split}
\end{equation}
If the term corresponding to the upper and lower bands is neglected, the gap equation~(\ref{eq:gap_1_T}) reduces to $\Delta_A = \frac{Un_{\phi}}{2}\tanh\frac{\beta\Delta_A}{2}$. This result has been used in the last equality of Eq.~(\ref{eq:sfw_flat}). Eq.~(\ref{eq:sfw_flat}) is consistent with the general result for the superfluid weight at finite temperature in the flat-band limit as provided in Ref.~\cite{Sebastiano:2015}. This is rather surprising since one assumption has been made in the derivation of this result in Ref.~\cite{Sebastiano:2015}, namely that the order parameters are all equal $\Delta_\alpha = \Delta$, but this condition is not satisfied in the case of the Lieb lattice where $\Delta_A = \Delta_C \neq \Delta_B$. This can be traced back to the fact that when the derivatives of Eq.~(\ref{eq:GDG}) are taken all the terms proportional to $\Delta_B$ drop out. Eq.~\eqref{eq:sfw_flat} can be extended away from half-filling by using the block structure of the Bogoliubov de-Gennes Hamiltonian~\eqref{eq:BdG} (see also Eq.~\eqref{eq:W_half}). The result is
\begin{equation}
\left.[D_{\rm s}]_{i,j}\right|_{\rm f.b.} = \frac{1}{\pi \hbar^2}\frac{\Delta_A^2}{E_0}\tanh\frac{\beta E_0}{2}\left.\mathcal{M}^{\rm R}_{ij}\right|_{\rm f.b.}\,,
\end{equation}
with the quasiparticle energy given by $E_{0,\vc{k}} = E_0 = \sqrt{\mu^2+\Delta_A^2}$. 

The staggered hopping parametrized by $\delta$ breaks the symmetry of the square lattice with respect to rotations by $\ang{90}$. This means that $\mathcal{M}_{ij}^{\rm R}|_{\rm f. b.}$ is not a diagonal matrix for $\delta \neq 0$, but has a nonzero off-diagonal component $\mathcal{M}_{xy}^{\rm R}|_{\rm f. b.}$ while the diagonal components are equal $\mathcal{M}_{xx}^{\rm R}|_{\rm f. b.} = \mathcal{M}_{yy}^{\rm R}|_{\rm f. b.}$. The components of $\mathcal{M}_{ij}^{\rm R}|_{\rm f. b.}$ are shown in Fig.~5 of the main text as a function of $\delta$. The off-diagonal elements  are finite for all $\delta$, while the diagonal ones have a logarithmic singularity for $\delta = 0$. This singularity is due to the fact that the periodic Bloch functions are nonanalytic functions of the wavevector at the band intersection $\vc{k}a = (\pi,\pi)^T$ for $\delta = 0$. This signals that other bands have to be included in order to compute the superfluid weight.  The various contributions to the superfluid weight are shown in Fig.~4(a)-(d) in the main text. 
\begin{figure}
\begin{center}
\includegraphics[scale=0.8]{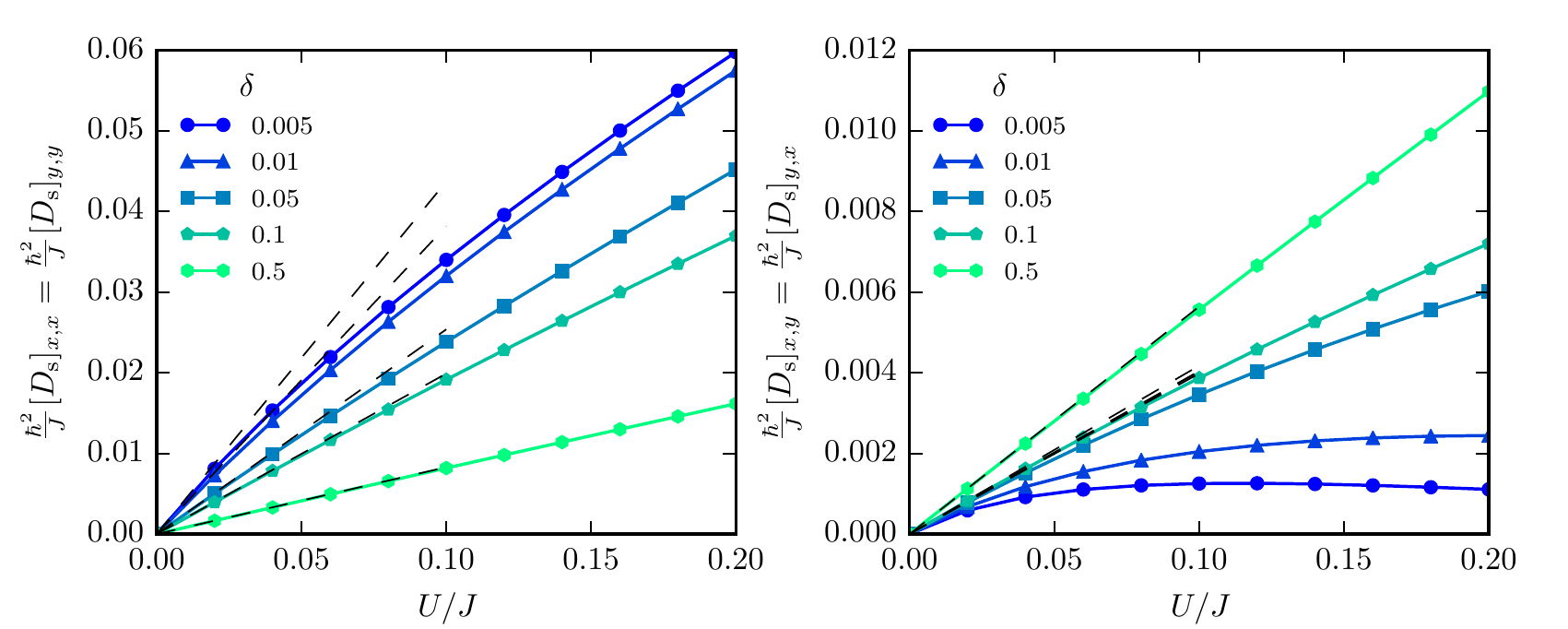}
\caption{\label{fig:Ds_vs_U} Total superfluid weight as function of $U/J$ for different values of the staggering parameter $\delta$ at zero temperature $T=0$. The diagonal components of the superfluid weight tensor are shown in the left panel while the off-diagonal ones are shown in the right panels. For large $\delta$ the superfluid weight is linear in $U$ since the flat band contribution is dominating. The black dashed line represents the flat band contribution Eq.~(\ref{eq:sfw_flat}). Deviations from linearity are more pronounced for small $\delta$ since the geometric contribution of the other bands plays an increasingly important role.  
} 
\end{center}
\end{figure}
On the other hand, in Fig.~\ref{fig:Ds_vs_U} we compare the total superfluid weight for different values of the staggering parameter $\delta$ at zero temperature. We find that for large $\delta$ the superfluid weight is linear in $U$, a fact that is explained by the dominant role of the flat band when the energy gap $E_{\rm gap}$ is much larger than $U$. Indeed Eq.~(\ref{eq:sfw_flat}) gives the slope of $D_{{\rm s}}$ around $U = 0$. On the contrary, for small $\delta$ pronounced deviations from linearity can be seen, an effect due to the other bands geometric contribution $\left.D_{\text{s,geom}}\right|_{\rm o.b.}$.  At $\delta = 0$ this implies that the superfluid weight has a diverging derivative at $U = 0$. However, the geometric contribution to the superfluid weight from the other band $D_{\rm s,geom}|_{\rm o.b.}$ ensures that the superfluid weight is finite even at $\delta = 0$. 

Note how the diagonal components $[D_{\rm s}]_{x,x} = [D_{\rm s}]_{y,y}$ are decreasing functions of $\delta$, while for $\delta = 0$ the off-diagonal elements are zero due to rotational symmetry, and their magnitude increases with $\delta$. Eventually, for $\delta = 1$ the superfluid weight tensor has a zero eigenvalue which implies that the superconducting state is unstable (see Fig.~5 in the main text). Indeed, long-range order cannot be established since the unit cells as defined in Fig.~1 in the main text are decoupled. However, the value of the order parameter $\Delta_A$ is essentially unaffected at the mean-field level when changing $\delta$ as shown in Fig.~3 in the main text. This unphysical behavior is due to the fact that BCS theory captures thermally excited quasiparticles, but not the thermal fluctuations of the order parameter phase or other collective modes. The phase fluctuations are  responsible for the collapse of the superconductive order with increasing $\delta$ and they are captured by Dynamical Mean Field Theory to some extent (see below).

\section{Appendix B: Exactness of the BCS wavefunction in the isolated flat-band limit}

In this section we prove that in case of a bipartite lattice that supports an isolated flat band ($U \ll E_{\rm gap}$) the BCS wavefunction  becomes an exact ground state when the flat band is partially filled. We start by considering a Hamiltonian with \textit{repulsive} Hubbard interaction that reads
\begin{equation}\label{eq:repulsive_hubbard}
\hat{\mathcal{H}} = \hat{\mathcal{H}}_{\text{kin}} + \hat{\mathcal{H}}_{\text{int}} \quad \text{with} \quad \mathcal{\hat{H}}_{\rm kin} = \sum_{\vc{k},\sigma}\hat{\vc{c}}_{\vc{k}\sigma}^\dagger H_{\vc{k}}\hat{\vc{c}}_{\vc{k}\sigma} \quad \text{and} \quad  \hat{\mathcal{H}}_{\text{int}} = U\sum_{\vc{i},\alpha}\hat{n}_{\vc{i}\alpha \uparrow}\hat{n}_{\vc{i}\alpha \downarrow},\quad U > 0\,.
\end{equation}
The vector $\vc{\hat{c}}_{\vc{k}\sigma}  = (\hat{c}_{A\vc{k}\sigma},\hat{c}_{B\vc{k}\sigma},\dots)^T$ collects the field operators $\hat{c}_{\alpha\vc{k}\sigma}$ relative to the orbitals $\alpha = A, B, \dots$ as defined in the main text. 
By definition bipartite lattices can be divided into two sublattices, $L1$ and $L2$, in such a way that the matrix elements of the single-particle kinetic Hamiltonian $H_{\vc{k}}$ between states belonging to the same sublattice are all zero. This means that $H_{\vc{k}}$ has the form
\begin{equation}\label{eq:chiral_form}
H_{\vc{k}} = 
\begin{pmatrix}
0 & A_{\vc{k}}^\dagger \\
A_{\vc{k}} & 0  
\end{pmatrix}\,.
\end{equation}
where $A_{\vc{k}}$ is an arbitrary rectangular matrix with the number of rows (columns) equal to the number of orbitals per unit cell in the $L1$ ($L2$) sublattice, given by  $|L1|/N_{\rm c}$ ($|L2|/N_{\rm c}$). Here $|L1|$ ($|L2|$) is the number of lattice sites in the $L1$ ($L2$) sublattice and $N_{\rm c}$ the number of unit cells.
The number of zero eigenvalues of a matrix $H_{\vc{k}}$ of the form given by Eq.~(\ref{eq:chiral_form}) is $N_{\rm f.b.} = \mathrm{dim}(H_{\vc{k}}) - \mathrm{rank}(A_{\vc{k}})- \mathrm{rank}(A_{\vc{k}}^\dagger) = \mathrm{dim}(H_{\vc{k}}) - 2\,\mathrm{rank}(A_{\vc{k}})$. Assuming $|L1|/N_{\rm c} \geq |L2|/N_{\rm c}$
one has in general $\mathrm{rank}(A_{\vc{k}}) = |L2|/N_{\rm c}$, therefore  $N_{\rm f.b.} = |L1|/N_{\rm c}-|L2|/N_{\rm c}$ is the number of flat bands with zero energy of the kinetic Hamiltonian. Due to the particle-hole symmetry of a Hamiltonian of the form~(\ref{eq:chiral_form}) the number of positive (negative) energy bands is given by $|L2|/N_{\rm c}$.
Specifically, in case of the Lieb lattice the $L1$ sublattice consists of the $A$ and $C$ sublattices and $L2$ is taken to be the $B$ sublattice, i.e. $|L1|/N_c = 2$ and $|L2|/N_c =1$ so that $N_{\textrm{f.b.}} =1$ as expected.

We further assume that the zero-energy flat bands are separated from the other bands by an energy gap $E_{\rm gap} \gg U$. At half-filling the total number of particles is $N_{\rm p} = (|L1|+|L2|)$. The negative energy bands are completely filled and accommodate $2|L2|$ particles. For small $U$ they can be neglected, therefore in the following we denote by $\ket{\emptyset}$ the state with the negative energy bands completely filled. At half-filling the remaining $N_{\rm c}N_{\rm f.b.}$ particles are accommodated in the zero-energy flat bands.
According to Lieb theorem~\cite{liebpaper}, the repulsive Hubbard model of Eq.~(\ref{eq:repulsive_hubbard}) on a bipartite lattice at half-filling has a ground state with total spin $S$ given by $2S = |L1| - |L2| = N_{\rm c}N_{\rm f.b.}$. The condition on the total spin implies that the particles in the flat bands can  be only in a completely polarized ferromagnetic state of the form $|\text{Ferro}\rangle = \prod_{\vc{k}} \Big(u \hat{d}^\dag_{0\vc{k}\downarrow} + v \hat{d}^\dag_{0\vc{k}\uparrow}\Big)|\emptyset\rangle$.
These states are degenerate due to spin rotational symmetry of Eq.~(\ref{eq:repulsive_hubbard}), indeed the parameters $u,v$ are normalized $|u|^2+|v|^2= 1$, but otherwise arbitrary.

The repulsive Hubbard model~\eqref{eq:repulsive_hubbard} on the Lieb lattice can be mapped to the attractive one by performing the following particle-hole transformation
\begin{align}
\label{phtrans}
&\hat{c}_{\vc{i}\uparrow \alpha} \rightarrow \hat{c}_{\vc{i}\uparrow \alpha} \nonumber \\
&\hat{c}_{\vc{i}\downarrow \alpha} \rightarrow s(\alpha) \hat{c}^\dag_{\vc{i}\downarrow \alpha},
\end{align}
where $s(\alpha) = 1$ for the lattice sites belonging to the $L1$ sublattice ($s(A) = s(C) = 1$) and $s(\alpha) = -1$ for the lattice sites belonging to the $L2$ sublattice ($s(B)  = -1$). To see the effect of this transformation, we expand the down-spin Bloch state operators  as follows:
\begin{align}
\label{expansion}
\hat{d}_{n\vc{k}\downarrow} = \frac{1}{\sqrt{N_{\rm c}}}\sum_{\vc{i},\alpha}e^{-i\vc{k}\cdot \vc{r}_{\vc{i}\alpha}} g^*_{n\vc{k}}(\alpha) \hat{c}_{\vc{i}\alpha\downarrow} \to \frac{1}{\sqrt{N_{\rm c}}}\sum_{\vc{i},\alpha}e^{-i\vc{k}\cdot \vc{r}_{\vc{i}\alpha}} g^*_{n\vc{k}}(\alpha) s(\alpha) \hat{c}^\dag_{\vc{i}\alpha\downarrow}\,.
\end{align}
By using Eq.~\eqref{eq:G_def} for $\gamma_A = \gamma_B$, Eq.~\eqref{expansion} and the fact that the flat band is supported only by the $A$ and $C$ sublattices (i.e. $g_{0\vc{k}}(B) = 0$), one can easily show that under the particle-hole transformation \eqref{phtrans} the operators $\hat{d}_{n\vc{k}\downarrow}$ transform as
\begin{align}
&\hat{d}_{0\vc{k}\downarrow} \rightarrow \hat{d}^\dag_{0(-\vc{k})\downarrow} \label{flatbandtrans} \\
&\hat{d}_{-(\vc{k})\downarrow} \rightarrow \hat{d}^\dag_{+(-\vc{k})\downarrow} \label{minustrans}\\
&\hat{d}_{+(\vc{k})\downarrow} \rightarrow \hat{d}^\dag_{-(-\vc{k})\downarrow}, \label{plustrans}
\end{align}
As a consequence the vacuum state transforms as $|\emptyset\rangle\rightarrow\prod_{\vc{k}} \hat{d}^\dag_{0(-\vc{k})\downarrow}|\emptyset\rangle$, while the ferromagnetic state becomes the BCS wavefunction
\begin{align}\label{eq:wave_BCS}
|\text{Ferro}\rangle &\to  \prod_{\vc{k}}\Big(u \hat{d}_{0(-\vc{k})\downarrow} + v \hat{d}^\dagger_{0\vc{k}\uparrow} \Big)\prod_{\vc{k}'}d^\dag_{0(-\vc{k}')\downarrow}|\emptyset\rangle = \prod_{\vc{k}}\Big[\Big(u \hat{d}_{0(-\vc{k})\downarrow} + v \hat{d}^\dagger_{0\vc{k}\uparrow} \Big)\hat{d}^\dag_{0(-\vc{k})\downarrow} \Big]|\emptyset\rangle \nonumber \\
&=  \prod_{\vc{k}}\Big(u + v \hat{d}^\dagger_{0\vc{k}\uparrow} \hat{d}^\dagger_{0(-\vc{k})\downarrow} \Big)|\emptyset\rangle = |\text{BCS}\rangle\,.\end{align}
Note that if we use the parametrization $u = \sqrt{1-\nu_{\rm f.b.}}$ and $v = e^{i\phi}\sqrt{\nu_{\rm f.b.}}$, then $\nu_{\rm f.b.} = \nu - 1$ is the flat-band filling and $e^{i\phi}$ is the arbitrary phase of the superconducting order parameter. Therefore the degeneracy of the ferromagnetic ground state translates into the degeneracy of the wavefunction~(\ref{eq:wave_BCS}) with respect to changes in the filling and in the superconducting order parameter phase. 
The result is that the BCS wavefunction is the exact ground state for an attractive Hubbard interaction at any fillings of the flat band. 
The proof can be extended to general bipartite lattices with $N_{\rm f.b.} \neq 0$. Indeed the particle-hole transformation in Eq.~\eqref{phtrans} is generic for single-particle Hamiltonians of the form~\eqref{eq:chiral_form}.

\section{Appendix C: Comparison between mean field BCS theory and dynamical mean field theory}

\begin{figure}
\begin{center}
\includegraphics[width=1.0\textwidth]{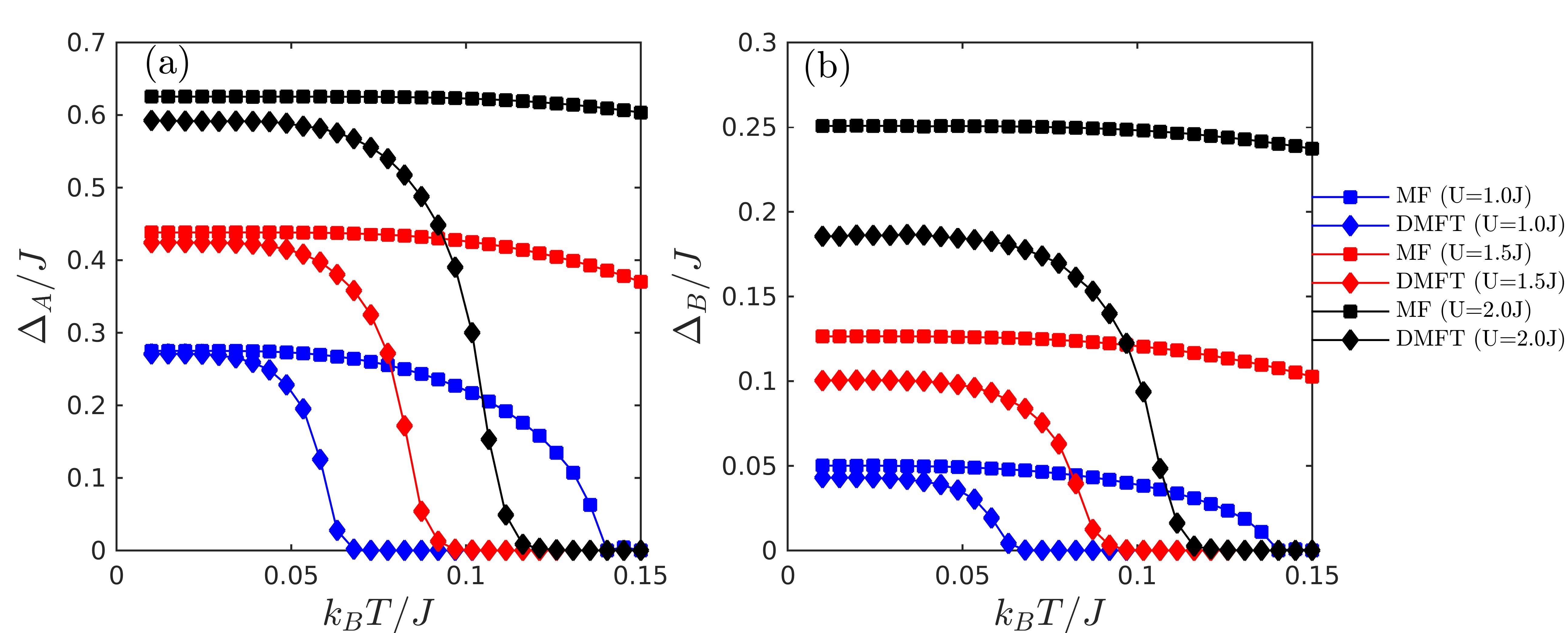}
\caption{\label{fig:mf_vs_dmft} Order parameters $\Delta_A$ (a) and $\Delta_B$ (b) for half-filled flat band computed by using DMFT and MF as a function of temperature $T$. The results are provided for three different interaction strengths. The data is for hopping coefficients without staggering ($\delta = 0$).} 
\end{center}
\end{figure}

\begin{figure}
\begin{center}
\includegraphics[width=1.0\textwidth]{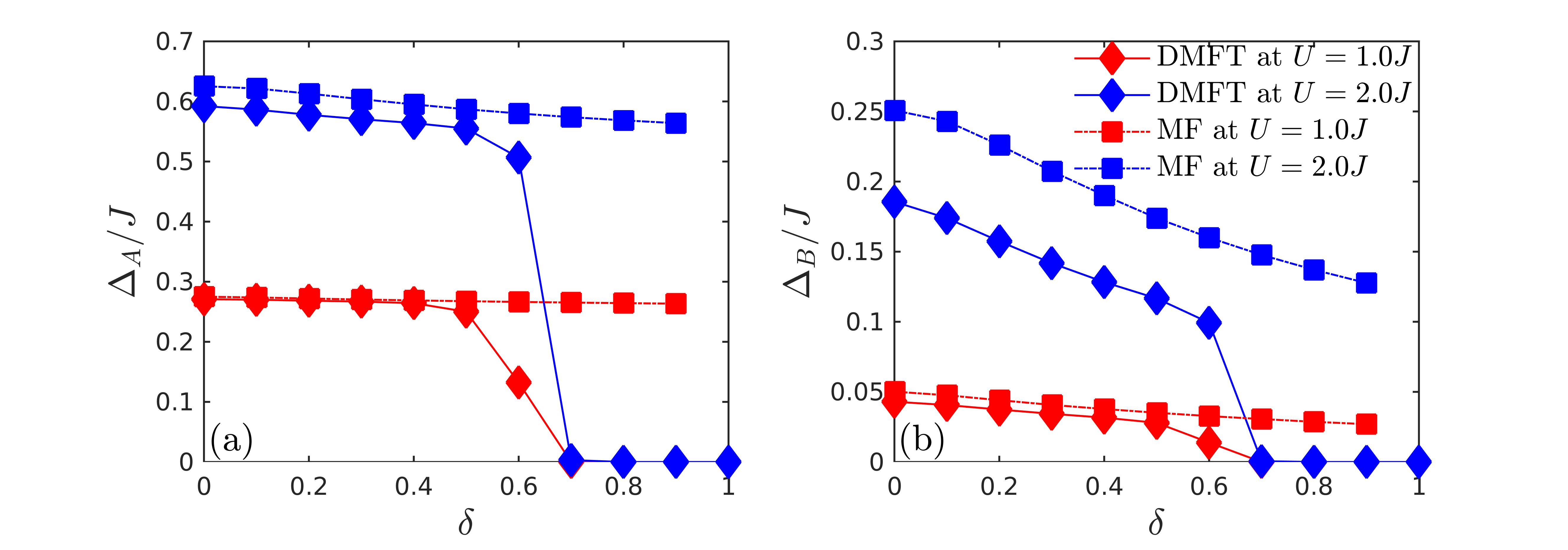}
\caption{\label{fig:mf_vs_dmft2} Order parameters $\Delta_A$ (a) and $\Delta_B$ (b) of half-filled flat band as a function of $\delta$ obtained by using MF (squares) and DMFT (diamonds) for two different interaction strengths ($U= 1.0J$ and $U=2.0J$). Here the temperature is set to $k_{\rm B}T=0.01J$.} 
\end{center}
\end{figure}

To check the validity of our BCS theory, we apply cellular dynamical mean-field theory \cite{RevModPhys.68.,RevModPhys.77.,PhysRevLett.87.186401} with the continuous-time interaction expansion (CT-INT) impurity solver \cite{ctint1,ctint2}. In our computations the impurity problem is chosen to consists of the three lattice sites within one unit cell which is then coupled self-consistently to the rest of the lattice. Inside the unit cell the correlations are treated exactly, whereas the coupling to the environment is treated at the mean-field level.

In Fig.~\ref{fig:mf_vs_dmft} we compare BCS with DMFT for half-filled flat band and three different interaction strengths $U= 1.0 J,\, 1.5 J,\, 2.0 J$ by presenting the order parameters $\Delta_A$ and $\Delta_B$ as a function of the temperature. We see that at high temperatures  BCS deviates notably from DMFT and overestimates the critical temperatures. Indeed, BCS neglects thermal fluctuations of the order parameter phase as discussed in Section~\ref{sec:sfw}, while they are included to a certain extent in DMFT.
On the other hand, at lower temperatures the agreement between the two methods is good, especially in case of $\Delta_A$. Because superfluidity in the flat band is related to a finite $\Delta_A$ rather than $\Delta_B$, we deduce that at low temperatures the BCS approach is reliable when investigating the superconductive properties of the flat band. 

We further compare the two methods in Fig.~\ref{fig:mf_vs_dmft2} where we plot $\Delta_A$ and $\Delta_B$ as a function of the staggering parameter $\delta$ obtained by BCS  and DMFT for two different interaction values, $U= 1.0J$ and $U=2.0J$.  This is the same plot as in Fig. 2 in the main text where $U = 0.4J$. One can see that, especially in case of $\Delta_A$, BCS is in good agreement with DMFT even for larger $U$.  Compared to the results in Fig. 2 of the main text, we also see that now the order parameter values computed by using DMFT are finite for larger staggering values. This is expected since the superfluid weight increases approximately linearly with the interaction strength and the system becomes correspondingly more robust against thermal fluctuations of the order parameter phase. 

\section{Appendix D: Exact-diagonalization calculation of Drude weight}

\begin{figure}
\begin{center}
\includegraphics[width=0.8\textwidth]{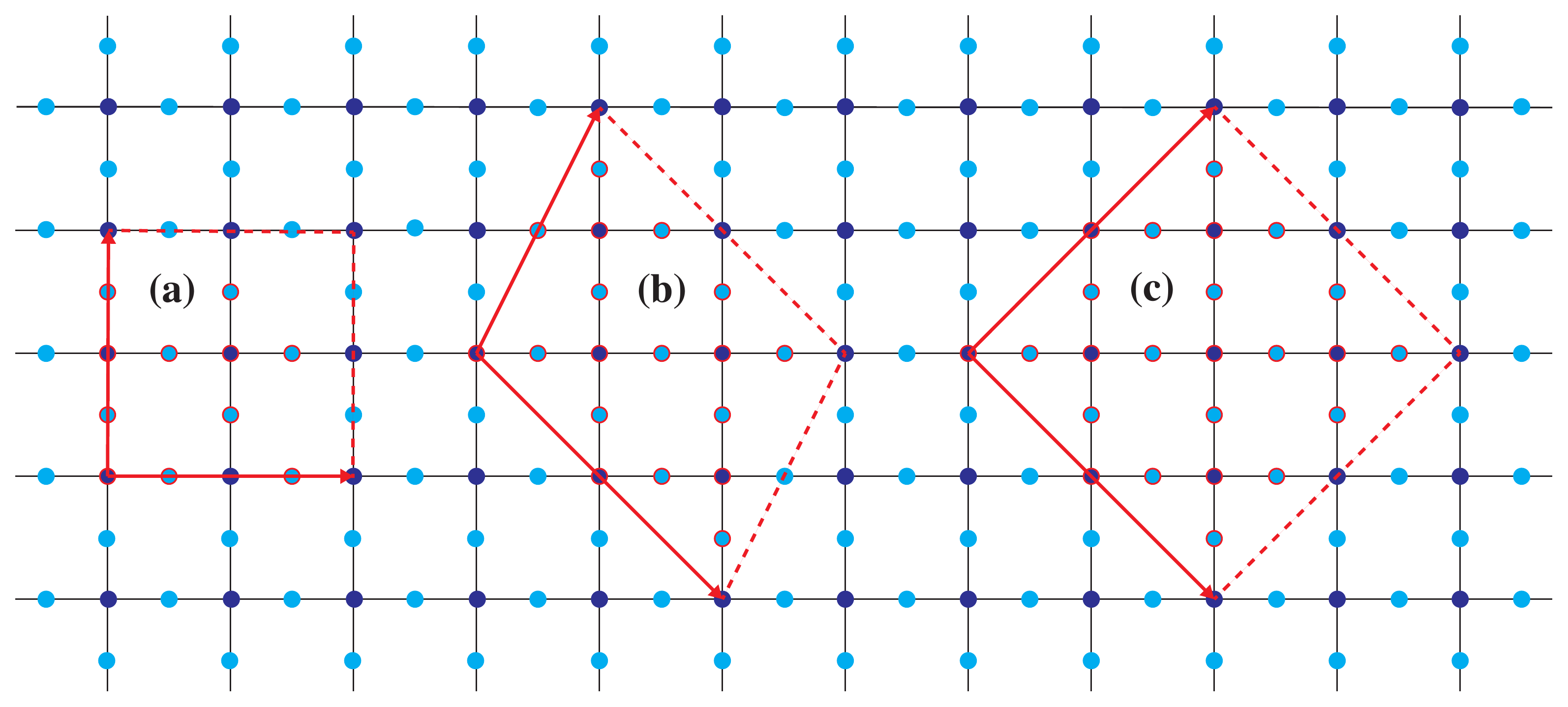}
\caption{\label{fig:lattice_for_ed}Finite-size cells used in the exact-diagonalization calculation. The cells include (a) 12, (b) 18, and (c) 24 sites marked by red circles. The arrows indicate lattice translation vectors for the periodic boundary conditions.}
\end{center}
\end{figure}

\begin{figure}
\begin{center}
\includegraphics[width=1.0\textwidth]{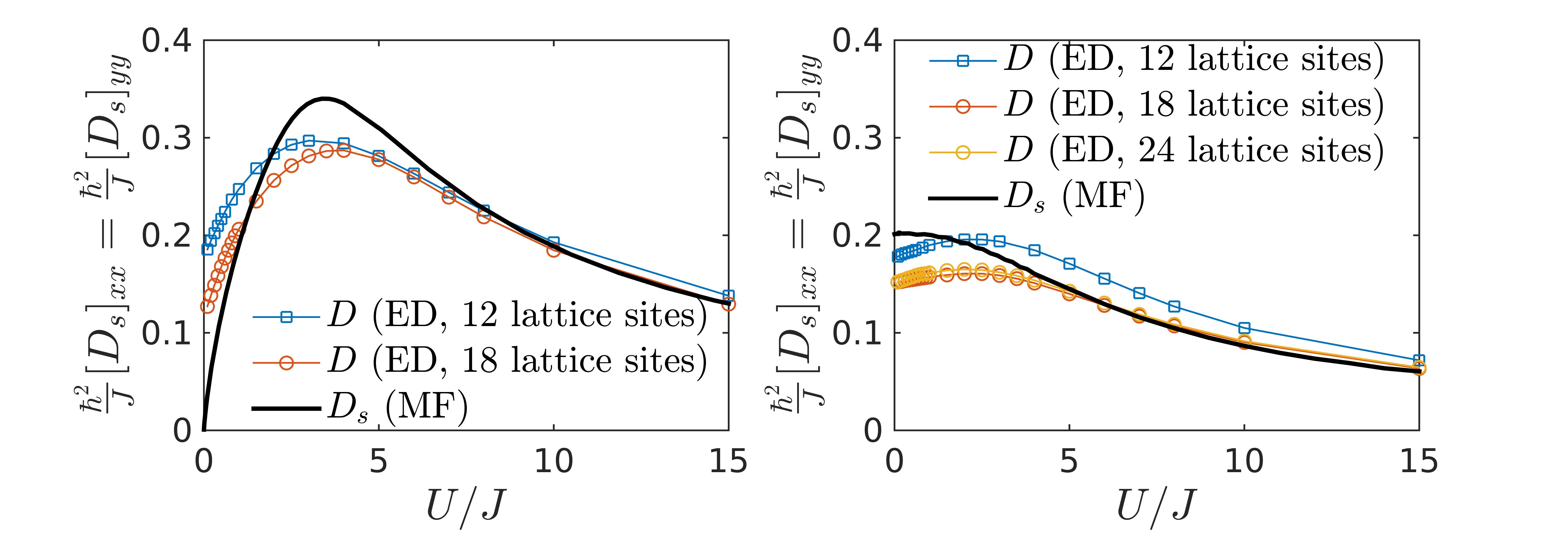}
\caption{\label{EDresults}Comparison between mean-field BCS and ED results at half-filling $\nu = 1.5$ (left) and half-filled upper band $\nu = 2.5$ (right). The data for the 24-sites cluster are essentially indistinguishable from the one relative to the 18 sites cluster.}
\end{center}
\end{figure}

The Drude weight is computed by employing the exact diagonalization (ED) method in the finite-size periodic Lieb cells of $12$, $18$, and $24$ lattice sites. The selected cell structures are shown in Fig.~\ref{fig:lattice_for_ed}. Following the standard procedures for the ED calculations (for instance, see \cite{Dagotto:1992}), the Drude weight in the $x$-direction is given as
\begin{equation}
[D]_{x,x} = -\frac{1}{V} \langle 0| \hat{K}_x |0 \rangle - \frac{2}{V} \sum_{n\neq 0} \frac{|\langle n| \hat{J}_x |0 \rangle |^2}{E_n - E_0}, 
\end{equation}
where the kinetic and current operators are defined as
\begin{eqnarray}
\hat{K}_x &=& -J\sum_{\mathbf{i},\sigma} (\hat c^\dagger_{\mathbf{i}A\sigma} \hat c_{\mathbf{i}B\sigma} + \hat c^\dagger_{\mathbf{i}A\sigma} \hat c_{\mathbf{i}B\sigma} )-J\sum_{\mathbf{i},\sigma}(\hat  c^\dagger_{\mathbf{i}B\sigma} \hat  c_{\mathbf{i}-\vc{\hat x},A\sigma} + \hat  c^\dagger_{\mathbf{i}-\vc{\hat x},A\sigma} \hat  c_{\mathbf{i}B\sigma} ), \\
\hat  J_x &=& iJ\sum_{\mathbf{i},\sigma} (\hat  c^\dagger_{\mathbf{i}A\sigma} \hat  c_{\mathbf{i}B\sigma} - \hat  c^\dagger_{\mathbf{i}A\sigma} \hat  c_{\mathbf{i}B\sigma} )+iJ\sum_{\mathbf{i},\sigma}(\hat  c^\dagger_{\mathbf{i}B\sigma} \hat  c_{\mathbf{i}-\vc{\hat x},A\sigma} - \hat  c^\dagger_{\mathbf{i}-\vc{\hat x},A\sigma} \hat  c_{\mathbf{i}B\sigma} ),
\end{eqnarray}
respectively, and are normalized by cell volume $V$. The Drude weight in the $y$-direction is defined in the same way by simply changing the unit vector connecting nearest-neighbor unit cells into $\hat{\mathbf{y}}$ and changing the orbital label $A\to C$. In our choices of the finite-size clusters, the computed values of $[D]_{x,x}$ and $[D]_{y,y}$ are numerically the same, and thus in the main text the Drude weight is denoted by $D$ without specifying a direction. The computation of the ground state energy $E_0$ and the ground state wavefunction $|0\rangle$ is done by using the Lanczos technique, and the second term of $D$ is evaluated through the continued fraction expansion of the regular part of the optical conductivity~\cite{Dagotto:1992}. The minimum computational memory requirement is 56 GB for half filling in the 18-site cluster and 175 TB for half filling in the 24-site cluster. All fillings are accessible in the 18-site cluster within our implementation of a parallel ED code, while $\nu=2.5$ is only treated in the 24-site cluster because of our limited computational resources. All the ED results for the fillings $\nu=1.5$ and $\nu=2.5$ are shown in Fig. \ref{EDresults} and compared with the BCS results. One can see that ED results converge when the cluster size is increased and are in good agreement with BCS results.

\section{Appendix E: Berezinsky-Kosterlitz-Thouless transition temperatures}
\begin{figure}
\begin{center}
\includegraphics[width=0.6\textwidth]{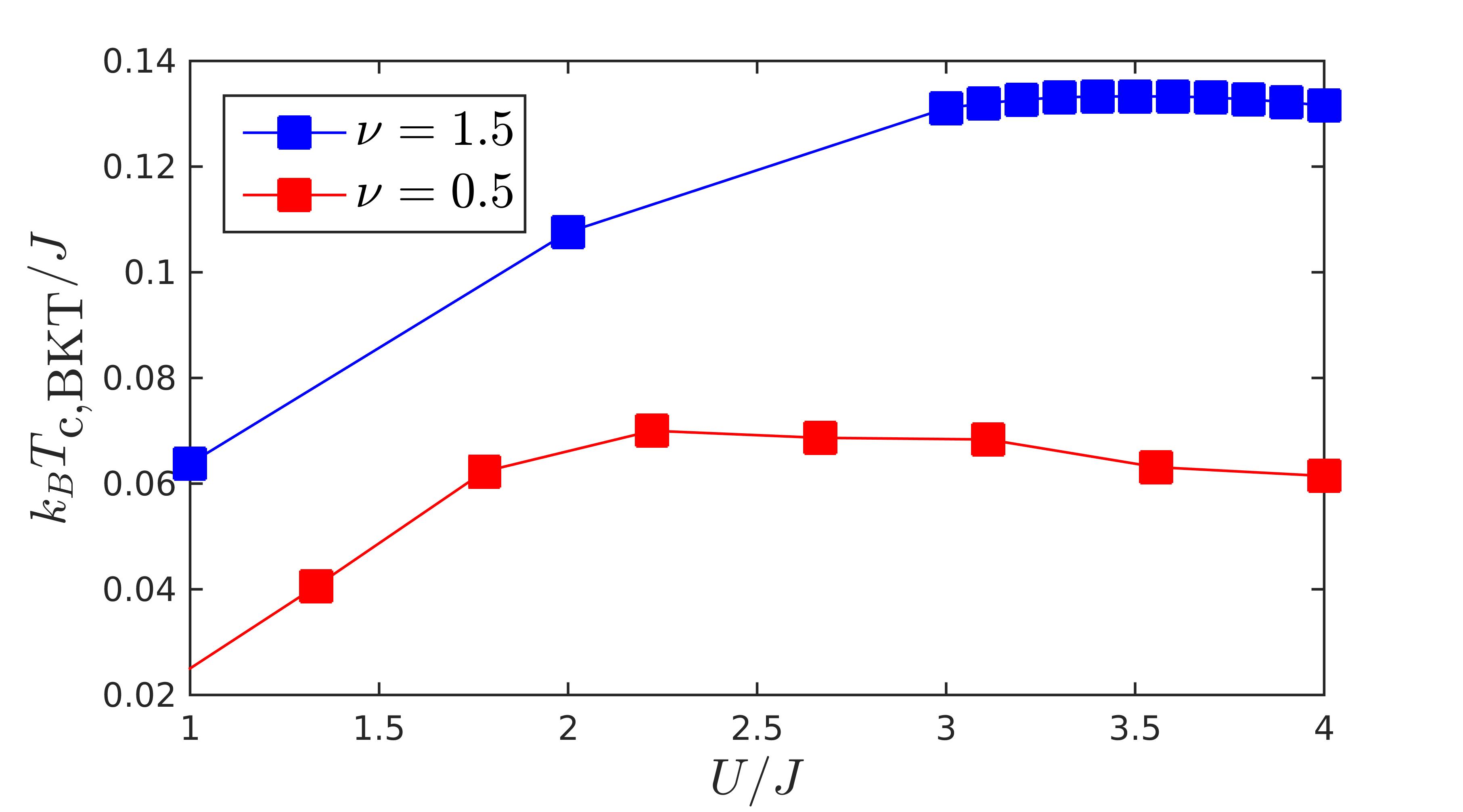}
\caption{\label{fig:tbkt}BKT transition temperatures for half-filled flat band ($\nu=1.5$) and half-filled lower dispersive band ($\nu=0.5$) as a function $U$. The staggering parameter is fixed to $\delta = 10^{-3}$.} 
\end{center}
\end{figure}

In two dimensions the Berezinsky-Kosterlitz-Thouless (BKT) transition temperature $T_{\text{c,BKT}}$ is defined by a well-known universal relation~\cite{bktpaper} that in our units reads
\begin{equation}
\frac{\hbar^2}{4} D_s(T_{\text{c,BKT}})  = \frac{2}{\pi}k_{\rm B} T_{\text{c,BKT}}.
\end{equation}
We use this formula to compute the transition temperature in our system, where the superfluid weight as a function of temperature $D_{\rm s}(T)$ is obtained from MF. In Fig.~\ref{fig:tbkt} we present $T_{\text{c,BKT}}$ as a function of $U$ for half filled flat band ($\nu=1.5$, blue curve) and for approximately half-filled lower dispersive band ($\nu\approx 0.5$, red curve) which is equivalent to half-filled upper dispersive band due to particle-hole symmetry of bipartite lattices. To compute the case $\nu=0.5$ one has to adjust the chemical potential $\mu$ for each value of $U$ in order to obtain the required filling for the dispersive band. This causes the small unphysical oscillations seen in the plot. One sees from Fig.~\ref{fig:tbkt} that the flat band yields higher transition temperatures by at least a factor of two in comparison with the dispersive bands. The transition temperature is maximized for the flat band around $U \approx 3.5J$ which yields the value $T_{\text{c,BKT}} \approx 0.133J$, whereas for the dispersive band the maximum occurs at $U\approx 2.2J$ with the value $T_{\text{c,BKT}} \approx 0.07J$. The maximum in the BKT critical temperature coincides approximately with the maximum in the superfluid weight (see Fig.~4(a) in the main text).

\bibliographystyle{apsrev4-1}

\end{document}